\documentclass{mn2e}
\usepackage[dvips]{graphicx}
\usepackage{psfig}
\usepackage{amssymb}

\title[Systematic uncertainties in cluster parameters based on broad-band
imaging]{Systematic uncertainties in the analysis of star cluster parameters
based on broad-band imaging observations}

\author[R. de Grijs et al.]{R. de Grijs,$^{1}$\thanks{E-mail:
R.deGrijs@sheffield.ac.uk} P. Anders,$^2$ H. J. G. L. M. Lamers,$^{3,4}$ N.
Bastian,$^{3,5}$ 
\newauthor U. Fritze--v. Alvensleben,$^2$
G. Parmentier,$^{6,7}$\thanks{Present address: Institute of Astronomy,
University of Cambridge, Madingley Road, Cambridge CB3 0HA} M. E.
Sharina,$^8$ and S. Yi$^9$
\\ 
$^1$ Department of Physics \& Astronomy, The University of Sheffield, Hicks
Building, Hounsfield Road, Sheffield S3 7RH \\
$^2$ Universit\"atssternwarte, University of G\"ottingen,
Geismarlandstr. 11, 37083 G\"ottingen, Germany \\
$^3$ Astronomical Institute, Utrecht University, Princetonplein 5,
3584 CC Utrecht, The Netherlands \\
$^4$ SRON Laboratory for Space Research, Sorbonnelaan 2, 3584 CA
Utrecht, The Netherlands \\
$^5$ European Southern Observatory, Karl-Schwarzschild-Strasse 2,
85748 Garching, Germany \\
$^6$ Institute of Astrophysics and Geophysics, Universit\'e de Li\`ege,
Sart-Tilman (B5c), 4000 Li\`ege, Belgium \\
$^7$ Astronomical Institute, University of Basel, Venusstrasse 7, CH-4102
Binningen, Switzerland \\
$^8$ Special Astrophysical Observatory, Russian Academy of Sciences, N.
Arkhyz, KChR, 369167, Russia \\
$^9$ Astrophysics, University of Oxford, Keble Road, Oxford OX1 3RH
}

\date{Received date; accepted date}
\pubyear{2005}

\begin{document}
\maketitle

\begin{abstract}
High-resolution {\sl Hubble Space Telescope (HST)} imaging
observations of star cluster systems provide a very interesting and
useful alternative for stellar population analyses to spectroscopic
studies with 8m-class telescopes.
%; the latter become prohibitive for large cluster samples.
Here, we assess the systematic uncertainties in (young) cluster age,
mass, and -- to a lesser extent -- extinction and metallicity
determinations, based on broad-band imaging observations with the {\sl
HST}. Our aim here is to intercompare the results obtained using a
variety of commonly used modelling techniques, specifically with
respect to our own extensively tested multi-dimensional approach. Any
significant differences among the resulting parameters are due to the
details of the various, independently developed modelling techniques
used, rather than to the stellar population models themselves. Despite
the model uncertainties and the selection effects inherent to most
methods used, we find that the {\it peaks} in the relative age and
mass distributions of a given young ($\lesssim 10^9$ yr) cluster
system can be derived relatively robustly and consistently, to
accuracies of $\sigma_t \equiv \Delta \langle \log( {\rm Age / yr} )
\rangle \le 0.35$ and $\sigma_M \equiv \Delta \langle \log( M_{\rm cl}
/ {\rm M}_\odot ) \rangle \le 0.14$, respectively, assuming Gaussian
distributions in cluster ages and masses for reasons of
simplicity. The peaks in the relative mass distributions can be
obtained with a higher degree of confidence than those in the relative
age distributions, as exemplified by the smaller spread among the peak
values of the mass distributions derived. This implies that mass
determinations are mostly insensitive to the approach adopted. We
reiterate that as extensive a wavelength coverage as possible is
required to obtain robust and internally consistent age and mass
estimates for the individual objects, with reasonable
uncertainties. Finally, we conclude that the {\it actual} filter
systems used for the observations should be used for constructing
model colours, instead of using conversion equations, to achieve more
accurate derivations of ages and masses.
\end{abstract}


\section{Introduction}
\label{intro.sec}

The {\it systematic} uncertainties in the determination of the ages,
masses, and -- to a lesser extent -- extinction and metallicity of
young extragalactic star clusters, based on broad-band imaging
observations, but using a variety of analysis approaches are poorly
understood. Our aim in this paper is to intercompare the results
obtained from {\sl Hubble Space Telescope (HST)} observations of
carefully selected young star cluster samples using a variety of
commonly used modelling techniques, and characterise the major
differences among the techniques. We will do this specifically with
respect to our own extensively tested multi-dimensional approach,
which we will use as our main benchmark.

\subsection{Star clusters as tracers of violent star forming episodes}

At one time or another during its lifetime, every galaxy will be
affected by the external gravitational effects, however minor, exerted
by its neighbour galaxies. Irrespective of the precise interaction
geometry, the results of such interactions are often striking:
depending on the available gas reservoir, violent star formation will
ensue, frequently predominantly in the guise of active star {\it
cluster} formation (e.g., Whitmore et al. 1999, de Grijs et al. 2001,
2003a,b, and references therein). Thus, where the above scenario
holds, the age distribution of a galaxy's star cluster population
reflects its violent interaction history.

The study of young and intermediate-age star cluster systems in a
variety of galactic environments out to $\lesssim 100$ Mpc has become
a major field in extragalactic astrophysics in recent years,
significantly stimulated by the superb imaging quality of the {\sl
HST}. One of the key diagnostic tools often utilised to predict the
fate of such cluster systems is the cluster luminosity function (CLF;
based on broad-band imaging observations).

Significant age spreads in young cluster systems -- which might still
be undergoing active cluster formation -- affect the observed CLF
(Meurer 1995, Fritze--v. Alvensleben 1998, 1999, de Grijs et al. 2001,
2003b). This might, in fact, make an intrinsically log-normal CLF
appear as a power-law (e.g., Miller et al. 1997,
Fritze--v. Alvensleben 1998); the exact shape of the intrinsic CLF,
whether a power law or a log-normal distribution, is still being
debated (e.g., Vesperini 2000, 2001, vs. Fall \& Zhang 2001; see also
Lamers et al. 2004). It is obviously very important to obtain accurate
age estimates for the individual clusters within such a system and to
correct the observed CLF to a common age, before interpreting it as an
{\it intrinsic} CLF (Fritze--v. Alvensleben 1999, de Grijs et
al. 2001, 2003b).

\subsection{Star cluster metallicities and the importance of interstellar 
dust} 

The metallicities of star clusters produced in the high-pressure
environments of galaxy interactions, mergers and starbursts are an
important discriminator against the old Milky Way-type globular
clusters (GCs), thought to be the oldest building blocks of most
nearby spiral and elliptical galaxies. They are expected to correspond
to the interstellar medium (ISM) abundances of the
interacting/starburst galaxies, and are therefore most likely
significantly more metal-rich than those of halo GCs in the Milky Way
and other galaxies with old GC systems. However, ISM abundances span a
considerable range among different galaxy types, from early-type
spirals to dwarf irregulars (e.g., Ferguson et al. 1998), and may also
exhibit significant radial gradients (Oey \& Kennicutt 1993, Zaritsky,
Kennicutt \& Huchra 1994, Richer \& McCall 1995). Hence, a
considerable metallicity range may be expected for star clusters
produced in interactions of various types of galaxies and even among
the clusters formed within one global galaxy-wide starburst.

A significant increase of the ISM abundance in massive gas-rich
galaxies may occur during strong bursts (Fritze--v. Alvensleben \&
Gerhardt 1994, their Fig. 12b). At the same time, typically within a
few $\times 10^8$ yr, some fraction of the gas enriched by dying
first-generation burst stars may well be shock-compressed to cool fast
enough to be built into later generations of stars or clusters
produced in the ongoing burst. The same effect may occur when multiple
bursts occur in a series of close encounters between two or more
galaxies before their final merger.

Precise (relative) metallicity determinations for individual young
star clusters are not only important to address these issues, but also
for the correct derivation of ages from broad-band colours or spectral
energy distributions (SEDs).

Dust extinction is often very important in young cluster systems. In
particular the youngest post-starburst galaxies and galaxies with
ongoing starbursts often show strong and patchy dust structures. For
instance, the youngest clusters in the overlap region of the two
galactic discs in the Antennae galaxies are highly obscured in the
optical and are best detected in near or mid-infrared observations
(Mirabel et al. 1998, Mengel et al. 2001). Similarly, Lan\c{c}on et
al. (2003) discovered one of the most massive young star clusters in
M82 based on near-infrared (NIR) {\sl HST} observations; at optical
wavelengths the cluster is completely obscured. Older merger remnants
like NGC 7252 or NGC 3921 seem to have blown their inner regions clear
of all the gas and dust left over from intense star formation (e.g.,
Schweizer et al. 1996). Extinction estimates toward individual
clusters are therefore as important as individual metallicity
estimates in order to obtain reliable ages and masses.

\subsection{Multi-passband photometry as a prime diagnostic}

Spectroscopy of individual massive young clusters, although feasible
today with 8m-class telescopes for the nearest systems, is very
time-consuming, since observations of large numbers of clusters are
required to obtain statistically significant results. Multi-passband
imaging is a very interesting and useful alternative, in particular if
it includes coverage of NIR and/or ultraviolet (UV) wavelengths (e.g.,
de Grijs et al. 2003c, Anders et al. 2004b). There are obviously
limitations to the accuracy of the cluster parameters derived from
broad-band imaging observations (e.g., de Grijs et al. 2003b,c, Anders
et al. 2004b, Bastian et al. 2004), but the {\it relative} overall
characteristics derived for the cluster {\it populations} as a whole
appear to be relatively robust.

In this paper we assess the systematic uncertainties in age and mass
determinations, and to a lesser extent also in extinction and
metallicity determinations, for young star cluster systems based on
the use of broad-band, integrated colours, employing a variety of
independently developed methods to analyse extragalactic star clusters
as so-called ``simple stellar populations'' (SSPs): star clusters are
the simplest objects to model, since they approximate single-age,
single-metallicity populations with a range of stellar
masses. Stochastic sampling effects of the stellar initial mass
function (IMF) also affect star cluster properties, in particular for
low-mass objects. However, since they affect broad-band photometry to
a smaller extent than spectroscopy, and because we are dealing here
with high-mass clusters only, we will not include these effects in
this paper. Our main aim in this paper is to intercompare the results
obtained for sets of well-calibrated cluster SEDs using a variety of
commonly used modelling techniques, specifically with respect to our
own extensively tested multi-dimensional approach (see Section
\ref{method5.sec}).

In order to determine the {\it absolute} systematic uncertainties
caused by the intrinsic differences in the models and methods in use
in the literature, we distributed sets of broad-band star cluster
photometry (described in Section \ref{data.sec}) to representatives of
the various groups active in this field, and requested them to derive
the cluster parameters using their specific methodology, wherever the
data allowed this to be done. The models and methods are described in
Section \ref{methods.sec}; we emphasize that most of the comparisons
among methods done in this paper should be considered relative to the
results obtained using the AnalySED method described in Section
\ref{method5.sec}. We summarise the results from applying our AnalySED
approach to a large grid of artificial cluster data in Section
\ref{artdata.sec}, in order to establish the theoretical benchmark for
further comparisons among approaches. In Section \ref{comparison.sec}
we compare the overall, relative parameter distributions, while in
Section \ref{detailed.sec} one-to-one comparisons between the various
method+model combinations for the individual clusters in both of our
samples are discussed; the results from each of the methods used for
both cluster samples are included in Appendix A. We extend this
discussion by considering the effects of converting the cluster
photometry to different filter systems (Section \ref{filters.sec}),
and conclude the paper in Section \ref{final.sec}.

\section{The cluster data sets}
\label{data.sec}

The field of stellar population modelling using extragalactic compact
star cluster systems has undergone a major expansion since
high-resolution, well-calibrated {\sl HST} observations became
available to the community. The application of stellar population
synthesis of galactic subcomponents has become almost trivial for
galaxies within $\sim 20-30$ Mpc, while star cluster population
modelling is very well feasible out to $\sim 100$ Mpc, at least for
the brighter (and therefore more massive) clusters within a given
cluster population (see, e.g., de Grijs et al. [2003d] and Pasquali,
de Grijs \& Gallagher [2003] for examples toward and close to the
distance limit).  Ongoing and future {\sl HST} programmes will
continue to provide high-resolution UV-optical-NIR imaging of large
samples of galaxies out to these distances. We therefore expect that
modelling simple stellar populations and their broad-band SEDs will
remain a key diagnostic tool for both the study of the evolution of
extragalactic star cluster systems and their relation to Milky
Way-type GCs, and for the analysis of galactic star formation and
interaction histories.

Therefore, we decided to focus our comparison of models and methods on
{\sl HST}-based imaging data. In addition, the calibration of {\sl
HST} measurements is well-understood and does, therefore, not
introduce additional uncertainties as caused by, e.g., temporal
variations in the atmospheric transmission that one would have to deal
with if ground-based observations were used.

We selected subsamples from large populations of young star cluster
systems extensively studied in the literature, which we required to be
among the brighter members of their respective populations (thus
minimising the observational uncertainties), as well as spanning a
large age range (based on preliminary analyses, as described
below). Ideally, we would have preferred to select cluster samples for
which both {\sl HST} measurements in a minimum of four broad-band
passbands could be obtained, as well as independently determined
parameters from spectroscopic observations. Unfortunately, such data
sets are yet not available, however. On the other hand, Schweizer,
Seitzer \& Brodie (2004) recently showed convincingly that
spectroscopic age determinations are not necessarily better or more
accurate than photometrically obtained ages, at least in the age range
from $\sim 100 - 500$ Myr.

Our basic cluster samples were taken from the following sources:

\begin{itemize}

\item NGC 3310, a nearby spiral galaxy exhibiting dominant star
cluster formation in a circumnuclear starburst ring. This galaxy was
covered by {\sl HST} by the largest possible wavelenth range. The full
set of eight broad-band {\sc stmag} magnitudes for the $\sim 300$
clusters in the galaxy's centre (located in the starburst ring and
outside of it), from F300W (``mid-UV'') to F205W were published and
analysed in de Grijs et al. (2003a,c). We selected 17 of these
clusters for the present analysis, all with well-determined,
high-quality photometry in the entire set of available passbands (see
Table \ref{ngc3310.tab}).

\begin{table*}
\caption[ ]{\label{ngc3310.tab}{\sl HST} {\sc stmag} photometry of the NGC 3310
cluster sample\\
The magnitudes are expressed in the {\sc stmag} {\sl HST} flight system,
derived from the count rates in the images as $m_{\sc stmag} = -2.5 \times
\log({\rm counts \; s}^{-1})$ + zero-point offset; the zero-point offset is
defined by the image header keywords {\sc photflam} and {\sc photzpt}.}
{\scriptsize
\begin{center}
\begin{tabular}{ccccccccc}
\hline
\hline
\multicolumn{1}{c}{ID} & \multicolumn{1}{c}{$m_{\rm F300W}$} &
\multicolumn{1}{c}{$m_{\rm F336W}$} & \multicolumn{1}{c}{$m_{\rm F439W}$} &
\multicolumn{1}{c}{$m_{\rm F606W}$} & \multicolumn{1}{c}{$m_{\rm F814W}$} &
\multicolumn{1}{c}{$m_{\rm F110W}$} & \multicolumn{1}{c}{$m_{\rm F160W}$} &
\multicolumn{1}{c}{$m_{\rm F205W}$} \\
\hline
G1-01 & $18.176 \pm 0.034$ & $18.207 \pm 0.038$ & $18.735 \pm 0.064$ & $19.105 \pm 0.046$ & $20.173 \pm 0.059$ & $21.162 \pm 0.064$ & $22.157 \pm 0.119$ & $22.451 \pm 0.063$ \\
G1-02 & $17.498 \pm 0.032$ & $17.445 \pm 0.027$ & $17.905 \pm 0.033$ & $18.938 \pm 0.062$ & $20.006 \pm 0.066$ & $20.858 \pm 0.042$ & $22.255 \pm 0.091$ & $22.691 \pm 0.089$ \\
G1-03 & $19.198 \pm 0.080$ & $19.352 \pm 0.077$ & $19.398 \pm 0.078$ & $20.439 \pm 0.095$ & $20.958 \pm 0.057$ & $21.482 \pm 0.049$ & $22.285 \pm 0.063$ & $23.089 \pm 0.098$ \\
G1-04 & $19.174 \pm 0.042$ & $19.500 \pm 0.095$ & $19.858 \pm 0.113$ & $20.349 \pm 0.085$ & $20.744 \pm 0.040$ & $21.231 \pm 0.050$ & $21.762 \pm 0.043$ & $22.412 \pm 0.034$ \\
G1-05 & $19.983 \pm 0.182$ & $20.121 \pm 0.181$ & $20.032 \pm 0.174$ & $20.593 \pm 0.121$ & $21.062 \pm 0.096$ & $21.740 \pm 0.079$ & $22.311 \pm 0.069$ & $23.377 \pm 0.079$ \\
G1-06 & $17.523 \pm 0.023$ & $17.643 \pm 0.022$ & $18.402 \pm 0.035$ & $18.796 \pm 0.040$ & $20.202 \pm 0.061$ & $21.239 \pm 0.102$ & $22.260 \pm 0.158$ & $22.448 \pm 0.173$ \\
G1-07 & $16.646 \pm 0.011$ & $16.788 \pm 0.011$ & $17.248 \pm 0.030$ & $17.411 \pm 0.030$ & $18.710 \pm 0.033$ & $19.607 \pm 0.038$ & $20.853 \pm 0.072$ & $20.870 \pm 0.060$ \\
G1-08 & $20.539 \pm 0.168$ & $20.887 \pm 0.233$ & $20.553 \pm 0.124$ & $21.124 \pm 0.089$ & $22.010 \pm 0.110$ & $22.668 \pm 0.134$ & $23.493 \pm 0.192$ & $25.675 \pm 1.261$ \\
G1-09 & $20.172 \pm 0.078$ & $19.857 \pm 0.066$ & $20.091 \pm 0.079$ & $20.896 \pm 0.089$ & $21.730 \pm 0.092$ & $22.065 \pm 0.141$ & $22.861 \pm 0.187$ & $22.946 \pm 0.160$ \\
G1-10 & $18.186 \pm 0.040$ & $18.182 \pm 0.038$ & $18.744 \pm 0.068$ & $19.127 \pm 0.052$ & $20.279 \pm 0.075$ & $20.925 \pm 0.049$ & $21.911 \pm 0.043$ & $22.463 \pm 0.086$ \\
G1-11 & $19.174 \pm 0.071$ & $19.305 \pm 0.059$ & $19.511 \pm 0.080$ & $20.259 \pm 0.094$ & $20.465 \pm 0.104$ & $21.390 \pm 0.160$ & $21.742 \pm 0.160$ & $22.770 \pm 0.188$ \\
G1-12 & $17.336 \pm 0.115$ & $17.467 \pm 0.118$ & $17.864 \pm 0.085$ & $18.714 \pm 0.162$ & $19.574 \pm 0.110$ & $20.386 \pm 0.116$ & $21.278 \pm 0.104$ & $21.716 \pm 0.111$ \\
G1-13 & $19.694 \pm 0.115$ & $20.563 \pm 0.234$ & $19.594 \pm 0.116$ & $21.057 \pm 0.199$ & $21.284 \pm 0.116$ & $21.629 \pm 0.098$ & $21.955 \pm 0.072$ & $22.847 \pm 0.107$ \\
G1-14 & $20.622 \pm 0.370$ & $20.219 \pm 0.219$ & $20.308 \pm 0.250$ & $21.032 \pm 0.175$ & $21.208 \pm 0.109$ & $21.774 \pm 0.105$ & $22.161 \pm 0.094$ & $23.289 \pm 0.217$ \\
G1-15 & $19.827 \pm 0.228$ & $19.870 \pm 0.174$ & $19.891 \pm 0.143$ & $20.384 \pm 0.071$ & $21.024 \pm 0.068$ & $21.604 \pm 0.131$ & $22.128 \pm 0.166$ & $23.244 \pm 0.268$ \\
G1-16 & $21.149 \pm 0.804$ & $21.063 \pm 0.600$ & $20.762 \pm 0.360$ & $20.902 \pm 0.158$ & $21.270 \pm 0.090$ & $21.876 \pm 0.081$ & $22.431 \pm 0.113$ & $22.880 \pm 0.110$ \\
G1-17 & $19.787 \pm 0.094$ & $19.967 \pm 0.121$ & $19.455 \pm 0.080$ & $20.818 \pm 0.117$ & $20.903 \pm 0.080$ & $21.218 \pm 0.056$ & $21.609 \pm 0.056$ & $22.234 \pm 0.059$ \\
\hline
\end{tabular}
\end{center}
}
\end{table*}

\item NGC 4038/39 (the ``Antennae''). Standard Johnson-Cousins {\it
UBVI} photometry and H$\alpha$ equivalent widths (EWs) for its large
population of young to intermediate-age star clusters were obtained
during a number of {\sl HST} imaging campaigns by Whitmore and
collaborators (see, e.g., Whitmore et al.  1999). Of the 20 objects
selected for the purpose of this paper (Table \ref{antennae.tab}), all
have well-determined {\it UBVI} magnitudes (obtained via conversion of
the {\sl HST} flight system magnitudes using the Holtzman et
al. [1995] conversion equations; but see Section \ref{filters.sec}),
while 10 of them have measured H$\alpha$ EWs as well.

\begin{table}
\caption[ ]{\label{antennae.tab}Johnson-Cousins photometry and H$\alpha$
EWs of the NGC 4038/39 cluster sample\\
The broad-band photometry is in magnitudes; $1\sigma$ photometric
uncertainties are on the order of 0.08 mag for all passbands. The H$\alpha$
EWs are expressed in {\AA}.}
{\scriptsize
\begin{center}
\begin{tabular}{cccccc}
\hline
\hline
\multicolumn{1}{c}{ID} & \multicolumn{1}{c}{\it U} & \multicolumn{1}{c}{\it B}
& \multicolumn{1}{c}{\it V} & \multicolumn{1}{c}{\it I} &
\multicolumn{1}{c}{log( EW$_{{\rm H}\alpha}$ )} \\
\hline
G2-01 & 21.944 & 21.779 & 21.550 & 21.098  &  $\cdots$ \\
G2-02 & 21.982 & 21.827 & 21.569 & 21.098  &  $\cdots$ \\
G2-03 & 23.118 & 23.075 & 22.602 & 21.694  &  $\cdots$ \\
G2-04 & 20.437 & 21.001 & 20.679 & 20.086  &  2.942 \\
G2-05 & 21.786 & 21.870 & 21.212 & 19.995  &  2.481 \\
G2-06 & 21.759 & 21.659 & 21.508 & 21.148  &  0.700 \\
G2-07 & 18.467 & 19.145 & 19.010 & 18.634  &  1.328 \\
G2-08 & 21.534 & 21.922 & 21.292 & 20.346  &  2.853 \\
G2-09 & 21.388 & 21.412 & 20.800 & 20.041  &  $\cdots$ \\
G2-10 & 20.273 & 20.732 & 20.296 & 19.516  &  1.350 \\
G2-11 & 23.861 & 23.707 & 22.769 & 21.540  &  $\cdots$ \\
G2-12 & 18.066 & 18.831 & 18.700 & 18.566  &  2.487 \\
G2-13 & 24.436 & 24.271 & 23.316 & 22.181  &  $\cdots$ \\
G2-14 & 23.377 & 23.673 & 22.261 & 21.157  &  3.728 \\
G2-15 & 18.557 & 19.428 & 19.064 & 18.919  &  3.497 \\
G2-16 & 20.159 & 20.511 & 20.321 & 19.823  &  0.348 \\
G2-17 & 19.420 & 19.966 & 19.656 & 18.928  &  $\cdots$ \\
G2-18 & 22.353 & 22.196 & 21.931 & 21.530  &  $\cdots$ \\
G2-19 & 23.285 & 23.332 & 22.544 & 21.511  &  $\cdots$ \\
G2-20 & 24.182 & 23.724 & 22.809 & 21.602  &  $\cdots$ \\
\hline
\end{tabular}
\end{center}
}
\end{table}

\end{itemize}

\section{Models and Methods}
\label{methods.sec}

Although the methods used to derive the global parameters of our cluster
samples each have their own merits and disadvantages, there is significant
overlap among both the extinction laws and the simple stellar population (SSP)
models used for the stellar synthesis modelling.

Therefore, we will first summarise the main characteristics of the SSP
models and extinction laws used in this project. Subsequently, in
Sections \ref{method1.sec}--\ref{method5.sec} each of the methods
employed to obtain the basic cluster parameters are described in
detail, roughly in order of increasing complexity and sophistication.

\subsection{Extinction laws}

In Sections \ref{method1.sec}--\ref{method5.sec} below, we will use a
variety of Galactic extinction laws, as published by Savage \& Mathis
(1979; Sections \ref{method2.sec} and \ref{method4.sec}), Rieke \&
Lebofsky (1985; Section \ref{method1.sec}), Voshchinnikov \& Il'in
(1987; Section \ref{method2.sec}), and Fitzpatrick (1999; Section
\ref{method3.sec}), as well as the starburst galaxy extinction law of
Calzetti et al. (1994; Section \ref{method5.sec}).

\begin{figure}
\psfig{figure=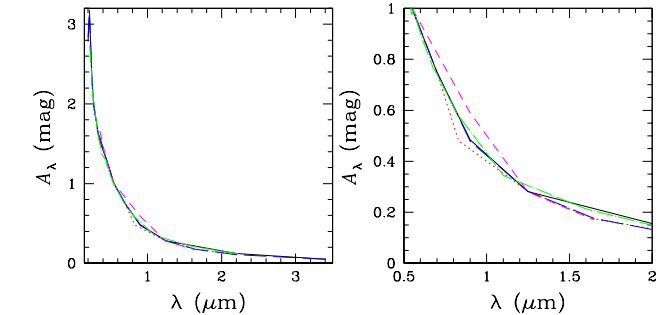,width=9cm}
\vspace{-4cm}
\caption[]{Comparison of the various extinction curves adopted; the
right-hand panel is a zoomed-in version of the left-hand
panel. Galactic extinction laws: solid lines -- Savage \& Mathis
(1979); dotted lines -- Rieke \& Lebofsky (1985); short-dashed lines
-- Voshchinnikov \& Il'in (1987); long-dashed lines -- Fitzpatrick
(1999). Starburst galaxy extinction law: dot/long-dashed lines --
Calzetti et al. (1994).}
\label{extcurves.fig}
\end{figure}

In the left-hand panel of Fig. \ref{extcurves.fig} we show these
extinction laws in relation to each other over the wavelength range of
interest for the present study, normalised at an extinction of 1 mag
in the {\it V} band at 5500{\AA}, $A_V = 1$ mag. In the right-hand
panel we zoom in to display the differences among the individual
extinction laws from 0.5 to 2.0 $\mu$m. From a comparison of the
individual extinction curves in the right-hand panel, it is clear that
the differences are generally $\lesssim 0.05$ mag at wavelengths
longward of $1 \mu$m and shortward of $\sim 0.8 \mu$m (with the
exception of the Voshchinnikov \& Il'in [1987] extinction law). In the
intermediate wavelength range, the differences are mainly driven by
the Rieke \& Lebofsky (1985) Galactic extinction law on the one hand
and the Voshchinnikov \& Il'in (1987) curve on the
other. Nevertheless, representative differences from the mean
generally do not exceed 0.1 mag, even at these wavelengths, and are
often significantly smaller.

If we place these differences among the extinction laws in their
proper context, i.e., in comparison with the observational data
presented in Section \ref{data.sec}, we see that they are generally of
the same order or smaller than the observational uncertainties. In
addition, as we will see in Section \ref{comparison.sec}, the vast
majority of the sample clusters are characterised by $A_V \ll 1$ mag,
so that the differences among the various extinction laws become
negligible for our sample clusters, irrespective of the analysis
approach adopted. This conclusion is further strengthened if we
realise that the wavelengths crucial for a successful determination
of, in particular, extinction values and metallicities are the bluest
optical/UV and the redder NIR passbands (see, e.g., de Grijs et
al. 2003c, Anders et al. 2004b), where the differences among the
individual extinction curves are smallest.

\subsection{Simple stellar population models}

All of the methods used in this study rely on a comparison of the
observational broad-band SEDs with a grid of model SEDs, in the sense
that the star clusters are assumed to represent ``simple'' stellar
populations, i.e., single-age, single-metallicity populations with a
range of stellar masses determined by a given stellar IMF.  The main
differences among the various SSPs used for the comparison done in
this paper are related to the use of different descriptions of the
input physics, such as (i) a variety of stellar tracks or isochrones,
which may or may not include critical phases in stellar evolution,
such as the red supergiant (RSG) phase, Wolf-Rayet stars, and the
thermally-pulsing AGB (TP-AGB) phase; (ii) slightly different IMF
descriptions; and (iii) different (or no) treatment of nebular
emission, particularly in the early phases of stellar evolution.

Each of the approaches used in this paper is based on a comparison
with a specific set of stellar evolutionary synthesis models. We will
use (i) the most commonly used set of models developed by Bruzual \&
Charlot (1993, 1996 [BC96], 2000 [BC00]; recently updated, 2003), (ii)
the Starburst99 models of Leitherer \& Heckman (1995) and Leitherer et
al. (1999), which are specifically matched to analyse the young
stellar populations in starburst and interacting galaxies, and (iii)
the G\"ottingen SSP models {\sc galev} (Kurth et al.  1999, Schulz et
al. 2002, Anders \& Fritze--v. Alvensleben 2003).

\subsubsection{The Bruzual \& Charlot SSP models}

The basic assumptions of the modern sets of the Bruzual \& Charlot SSP
models (used in Sections \ref{method1.sec}, \ref{method3.sec} and
\ref{method4.sec}), were first developed in Charlot \& Bruzual
(1991). The versions of the code used in this paper, BC96 and BC00,
include a description of SSP evolution for a range of metallicities
from $Z=0.0004$ to $Z = 0.10$.
%(with intermediate metallicities of $Z=0.004, 0.008, 0.02 = Z_\odot,$ 
%and 0.05)
The BC96 models and more recent versions are (mostly) based on the
evolutionary tracks of the Padova group (Bressan et al. 1993, Fagotto
et al. 1994a,b,c; with additional empirical spectra for stellar masses
between 0.1 and 0.7 M$_\odot$; Charlot \& Bruzual 1991), and cover an
age range from $1.25 \times 10^5$ to $2 \times 10^{10}$ yr, typically
computed using 220 unequally spaced age intervals. The important
TP-AGB phase (see Section \ref{galev.sec}) is treated
semi-empirically. These models cover all of the important phases of
stellar evolution from the zero-age main sequence to the post-AGB
phase and beyond, for stars with effective temperatures of $2000 \le
T_{\rm eff} ({\rm K}) \le 50,000$. The stellar spectra are based on
the theoretical spectral library compiled by Lejeune, Cuisinier \&
Buser (1997, 1998), which use the theoretical stellar atmosphere
calculations of Kurucz (1979), Fluks et al. (1994), and Bessell et
al. (1989, 1991). Lejeune et al. (1997, 1998) have corrected the
stellar continuum shapes from the Kurucz (1979) models to agree with
observed colours from the UV to the {\it K} band. 
%The most recently updated models, BC03, include the library of
%observed stellar spectra of Le Borgne et al. (2003), covering the
%wavelength range $3200 < \lambda < 9500${\AA}.
SSPs, covering the wavelength range from the extreme UV (5{\AA}) to
the far-infrared (100$\mu$m) with a resolution depending on the
spectral range, were calculated assuming a Salpeter (1955)-type IMF,
$\xi(m) \propto m^{-\alpha}$ with $\alpha = 2.35$ and masses ranging
from $\sim 0.1 {\rm M}_\odot$ up to 125 M$_\odot$.

\subsubsection{The Starburst99 models}
\label{sb99.sec}

The Leitherer et al. (1999) Starburst99 models (used in Sections
\ref{method2.sec} and \ref{method4.sec}) constitute an improved and
extended version of the suite of models initially published by
Leitherer \& Heckman (1995). These models were specifically developed
for the evolutionary synthesis analysis of populations of massive
stars, and are best suited to the conditions typically found in
starburst environments. They are based on the Geneva stellar evolution
models and the new model atmosphere grid compiled by Lejeune et
al. (1997). The tracks of Meynet et al. (1994) were used for stars
with masses in excess of 12--25 M$_\odot$ (depending on metallicity),
with the enhanced mass loss prescription in order to better
approximate most Wolf-Rayet properties (except the Wolf-Rayet
mass-loss rate itself) compared to the standard mass loss
scenario. For stars with masses in the range $0.8 \le m_\ast / {\rm
M}_\odot \le 12$, they used the standard mass-loss tracks of Schaller
et al. (1992), Schaerer et al. (1993a,b) and Charbonnel et
al. (1993). These tracks include the early AGB evolution until the
first thermal pulse for stars with masses $m_\ast > 1.7 {\rm
M}_\odot$. The Starburst99 models also include observational
high-resolution UV spectra, to allow for the analysis of stellar and
interstellar absorption lines and line profiles at various
metallicities.

The SSP models cover an age range between $10^6$ and $10^9$ yr, with
an age resolution of 0.1 Myr, for all five metallicities, $Z = 0.001,
0.004, 0.008, 0.02,$ and 0.04 over the entire spectral range from the
extreme UV to the infrared. Nebular continuum emission is included in
the models in a simplified fashion; its contribution becomes important
when hot stars providing ionising photons (and thus line emission) are
present (see Section \ref{galev.sec}).

The synthesised models that we use in Section \ref{method4.sec} below were
calculated using a standard Salpeter-type IMF, characterised by stellar masses
in the range $1 \le m_\ast / {\rm M}_\odot \le 100$.
%the upper mass limit was chosen based on the observational evidence
%that the most massive stars have masses around 100 M$_\odot$ (e.g.,
%Kudritzki et al. 1991).

\subsubsection{The {\sc galev} G\"ottingen SSP models}
\label{galev.sec}

The {\sc galev} SSPs (used in Sections \ref{method2.sec},
\ref{method4.sec} and \ref{method5.sec}) are based on the set of
stellar evolutionary tracks (Kurth et al. 1999), and in later versions
the isochrones (Schulz et al. 2002, Anders \& Fritze--v. Alvensleben
2003), of the Padova group (with the most recent versions using the
updated Bertelli et al. [1994; and unpublished] isochrones; the latter
also include the TP-AGB phase) for the metallicity range of $0.0001
\le Z \le 0.05$, tabulated as five discrete metallicities ($Z =
0.0004, 0.004, 0.008, 0.02,$ and 0.05, corresponding to [M/H]
$\approx$ [Fe/H] = $-1.7, -0.7, -0.4, 0.0,$ and +0.4,
respectively). For lower-mass stars ($0.08 \le m_\ast / {\rm M}_\odot
\le 0.5$), which contribute very little to the integrated light of
young and intermediate-age SSPs governed by any standard,
Salpeter-type IMF, the Padova models are supplemented with the
Chabrier \& Baraffe (1997) theoretical calculations that include a new
description of stellar interiors of low-mass objects and use non-grey
atmosphere models.

The {\sc galev} models are furthermore once again based on the
theoretical stellar libraries of Lejeune et al. (1997, 1998) for a
broad range of metallicities. For stars hotter than $T_{\rm eff} =
50,000$ K, pure black-body spectra are adopted, as for the Bruzual \&
Charlot models. The full set of models spans the wavelength range from
90{\AA} to 160$\mu$m.

The Salpeter-type IMF assumed is characterised by a lower cut-off mass
of $0.15 {\rm M}_\odot$; the upper-mass cut-off ranges between 50 and
70 M$_\odot$, and is determined by the mass coverage of the Padova
isochrones for a given metallicity.

Kurth et al. (1999) cover ages between $1 \times 10^7$ to $1.6 \times
10^{10}$ yr, with an age resolution of $10^7, 10^8,$ and $10^9$ yr,
for ages $\le 10^8$, between $10^8$ and $10^9$ and $\ge 10^9$ yr,
respectively. Schulz et al. (2002) and Anders \& Fritze--v.
Alvensleben (2003) extended the age range down to $4 \times 10^6$ yr
(and slightly reduced the upper age limit to 14 Gyr), while improving
the age resolution to 4 Myr for ages up to 2.35 Gyr, and 20 Myr for
greater ages.

The Schulz et al. (2002) version includes important improvements with
respect to the older versions; they use the newer Padova isochrones
which include the important stellar evolutionary TP-AGB phase. At ages
ranging from $\sim 100$ Myr to $\sim 1$ Gyr, TP-AGB stars account for
25 to 40 per cent of the bolometric light, and for 50 to 60 per cent
of the {\it K}-band emission of SSPs (see Charlot 1996, Schulz et
al. 2002). Schulz et al. (2002) show that the effect of including the
TP-AGB phase results in redder colours for SSPs with ages between
$\sim 10^8$ and $10^9$ yr, with the strongest effect (up to $\gtrsim
1$ mag) being seen in $(V-K)$ for solar metallicity, and in $(V-I)$
for $Z \ge 0.5 Z_\odot$. Shorter-wavelength colours and lower
metallicity SSPs are less affected. Since most young to
intermediate-age star cluster systems observed in {\sl HST} passbands
equivalent to the standard $V$ and $I$ filters, are in fact aged
between about 100 Myr and 1 Gyr, and have often close-to solar
metallicities, inclusion of the TP-AGB phase in the models is
obviously important.

Finally, Anders \& Fritze--v. Alvensleben (2003) included gaseous
continuum emission and an exhaustive set of nebular emission lines to
the {\sc galev} suite, assuming comparable metallicities for the star
cluster and the surrounding ionised gas. Nebular emission is shown to
be an important contributor to broad-band fluxes during the first few
$\times 10^7$ yr of SSP evolution, the exact details depending on the
metallicity.

\subsubsection{Model comparison}
\label{models.sec}

First, we present a basic comparison among the SSP models used in this
paper. Kurth et al. (1999) and Schulz et al. (2002) concluded that,
compared to the models of BC96 and Bruzual \& Charlot (1993),
respectively, their sets of {\sc galev} models agree very well for
solar metallicity and a Salpeter IMF, for $(B-V)$ colours, and from
the UV up to $\sim 7000${\AA}, respectively. However, between
7000{\AA} and 12,000{\AA}, as well as in the NIR {\it H} and {\it K}
band regime, the BC96 flux contribution is considerably lower than
that of the Schulz et al. (2002) spectrum at those wavelengths, which
they attribute to the different treatment of the TP-AGB evolutionary
phase (with $T_{\rm eff} \sim 3000$ K).

Anders \& Fritze--v. Alvensleben (2003) compare the most up-to-date
{\sc galev} models that include nebular line and continuum emission
with the Starburst99 models. Despite the differences in the input
physics and the different sets of stellar tracks and/or isochrones
used by these teams, they conclude that the differences between both
sets of models are minor at short optical wavelengths (e.g., $(B-V)$
colours) during the first Gyr of evolution, and are mainly due to the
better time resolution of the Starburst99 models. Longer-wavelength
comparisons show larger differences, due to the different input
physics, and in particular a different treatment of RSGs and the
TP-AGB phase.

In summary, it appears that over most of the optical wavelength range
all of the commonly used SSP models are fairly similar, with minor
differences depending on the detailed input physics and the treatment
of the various evolutionary phases. At longer (NIR) and shorter
(bluer) wavelengths, the differences become more significant, and will
lead to systematic differences in the determination of the basic
properties of SSPs, as we will see below.

Quantitatively, for a given set of input physics, varying parameters
including the IMF slope, mass loss and convection prescriptions, one
can justify differences of up to $\sim 0.05$ mag in $(B-V)$ (e.g., Yi
2003). However, the difference between -- for instance -- the Padova
and Geneva stellar evolutionary tracks when used by an otherwise
identical SSP code amounts to $\Delta(B-V) \sim 0.05$ mag, and even
more in $\Delta(V-I)$ and $\Delta(V-K)$ (e.g., Leitherer et al. 1996,
Schulz et al. 2002).

It is clear from the outset that the application of the various SSP
models to our sets of sample clusters will result in significantly
different masses and mass distributions, simply because of the
different low and high-mass boundaries adopted for the Salpeter-type
IMF. The mass ratios expected to result from the Starburst99 : the
Bruzual \& Charlot : the {\sc galev} SSPs are 1 : 22.4 : 33.1, or in
logarithmic mass units, mass estimates based on the Bruzual \& Charlot
SSPs will result in masses that are 1.35 dex higher than those from
the Starburst99 models; the {\sc galev} masses will be 1.52 dex
more massive than the Starburst99 ones.

\subsection{Method 1: Optical/NIR sequential analysis (``Sequential
O/IR'')}
\label{method1.sec}

The Sequential O/IR method is a two-step approach to derive the age,
metallicity and extinction values associated with a given
cluster. First, the extinction is estimated using the $BVI$ passband
combination; subsequently, the extinction-corrected, intrinsic colours
are compared, in a least-squares sense, to the BC96 SSP models in
order to estimate the cluster age.

While the $(B-V)$ vs. $(V-I)$ colour-colour diagram is affected by the
well-known age-metallicity degeneracy, the age and extinction
trajectories are not entirely degenerate for this particular choice of
optical colours. For SSPs older than $\sim 100$ Myr (i.e., $(B-V)_0
\gtrsim 0.18$ mag), all age trajectories show the same, roughly linear
growth of the $(V-I)_0$ vs $(B-V)_0$ colours, irrespective of their
metallicity. As a consequence, for such ages, $(B-V)$ vs. $(V-I)$ SSP
analysis enables us to derive the visual extinction (and therefore the
intrinsic colours and magnitudes), prior to any age and metallicity
estimates (cf. de Grijs et al. 2001).

Using the intrinsic colours we can now derive the most representative
age (and metallicity), by minimising (in a least-squares sense),
\begin{equation}
\chi^2_{\rm min}(t,Z) = {\rm min} \left\{\sum_{i=1}^3
\left(\frac{{{\rm CI}_i}^{\rm intr} - {{\rm CI}_i}^{\rm
SSP}(t,Z)}{\sigma_{{\rm CI}_i}}\right)^2\right\}\,,
\end{equation}
where CI$_{i}^{\rm intr}$ and CI$_{i}^{\rm SSP}(t,Z)$ are the
intrinsic and the model-predicted colour indices in a given colour
denoted by $i$, respectively, for SSPs with age $t$ and metallicity
$Z$; $\sigma _{{\rm CI}_{i}}$ are the 1$\sigma$ uncertainties.

\begin{figure}
\psfig{figure=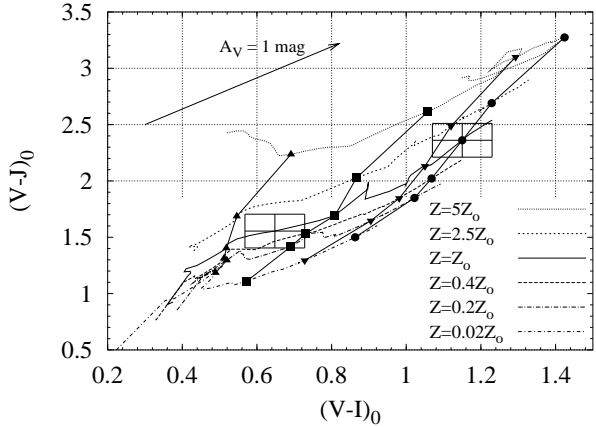,width=8.5cm}
\caption[]{$(V-I)_0$ vs. $(V-J)_0$ colour-colour diagram of the BC96
SSPs.  The thin curves represent isometallicity tracks; the thick
curves connecting filled symbols represent isochrones (from left to
right, log~$t/{\rm yr}$\,=\,8.5, 9.0, 9.5 and 10.0, respectively). As
an example, the thin boxes represent the medians of typical 1$\sigma$
error distributions for SSPs of solar metallicity at ages of 600\,Myr
and 10\,Gyr.  The arrow indicates the effects of reddening on these
models, for a visual extinction $A_V = 1\,{\rm mag}$.  For the sake of
clarity, only SSPs older than 100\,Myr are shown.}
\label{VIJ.fig}
\end{figure}

If NIR photometry is available, the cluster age {\it and} metallicity
can be derived simultaneously: for instance, the isochrones and
iso-metallicity tracks define a grid in $(V-I)$ vs. $(V-J)$ space,
thus allowing one to lift the age-metallicity degeneracy, as shown in
Fig. \ref{VIJ.fig}. Finally, cluster masses are obtained from their
luminosities via the age and metallicity dependent mass-to-light
ratio.

Our recent study of the intermediate-age star cluster population in
M82's region B provides a good example of what can be achieved if both
optical {\it and} NIR data are available. In that case, both the
age-extinction and the age-metallicity degeneracies can be
lifted. Further details, and a discussion about how photometric errors
propagate into extinction, age, metallicity and mass uncertainties are
given in Parmentier, de Grijs \& Gilmore (2003).

\subsection{Method 2: Reddening-free {\it Q} parameter analysis (``{\it
Q-Q}'')}
\label{method2.sec}

The basic {\it Q}-parameter analysis is a powerful method to determine
SSP ages and extinction values independently. The host galaxy's
internal extinction to a given cluster is derived from the $Q_{UBV}$
parameter (Johnson \& Morgan 1953),
\begin{eqnarray}
Q_{UBV} &=& (U-B)_0 - [{\rm E}(U-B)/{\rm E}(B-V)] \times (B-V)_0 \nonumber \\
&=& (U-B) - 0.72 \times (B-V),
\end{eqnarray}
To estimate the internal reddening we assess the loci of the clusters
in the $(B-V)$ vs $Q_{UBV}$ plane, compared to the intrinsic $(B-V)_0$
colours of the Starburst99 and {\sc galev} SSPs.

Subsequently, age estimates are obtained by minimising the clusters'
loci in the plane of the reddening-free $Q_{UBV}$ and $Q_{UBVI}$
parameters, where the latter is defined as (Whitmore et al. 1999)
\begin{eqnarray}
Q_{UBVI} &=& (U-B)_0 - [{\rm E}(U-B)/{\rm E}(V-I)] \times (V-I)_0 \nonumber \\
&=& (U-B) - 0.58 \times (V-I),
\end{eqnarray}
with respect to the Starburst99 models. Unfortunately, due to a loop
of the evolutionary tracks in {\it Q--Q} space, one cannot achieve
accurate age estimates in the range of log(Age/yr) from $\sim 6.5$ to
7.2 (see Whitmore et al. 1999). The availability of H$\alpha$
observations will greatly facilitate our age estimates in this age
range.

\subsection{Method 3: H$\alpha$ luminosities in addition to broad-band 
fluxes (``BB+H$\alpha$'')}
\label{method3.sec}

For those clusters for which we have H$\alpha$ flux or EW information
available, we compared the five observed magnitudes
($UBVI\mbox{H}\alpha$) with the predicted SEDs from the BC00 SSPs at
solar metallicity. For each available model age, we varied the
reddening between E$(B-V) = 0.0$ and 3.0 in steps of 0.02, and assume
the Galactic extinction curve of Fitzpatrick (1999). Each
model-age/reddening combination is scaled to match the cluster
$V$-band magnitude, and then compared with the observations using a
standard $\chi^2$ minimisation technique:
\begin{equation}
\chi^2_{\rm min}\bigl(t,{\rm E}(B-V)\bigr) = {\rm min}\Bigl\{
\sum_{\lambda} w_{\lambda}~(m_{\lambda}^{\rm model} - m_{\lambda}^{\rm
obs})^2 \Bigr\},
\end{equation}
where $\lambda=U,B,V,I,\mbox{H}\alpha$ and $w_{\lambda} = [(0.05)^2 +
\sigma_{m_{\lambda}}^{2}]^{-1}$.  Here, $\sigma_{m_{\lambda}}$ is the
photometric uncertainty (in magnitudes) for a given bandpass; we have
included an additional uncertainty of 0.05 mag, which represents the
uncertainties in the models themselves (Section \ref{models.sec}; see
Fall, Chandar \& Whitmore [2004] for validation of the method). Since
bright objects can have very small photometric uncertainties, this
keeps the weights from ``blowing up'' in the very bright object
regime.

The predicted H$\alpha$ model flux is calculated from the total number
of ionising photons under the assumption of photon-limited case B
recombination. When converting to magnitudes, we determined the
zero-point offset between the models and observations empirically, by
comparing the observed difference in ($m_{{\rm H}\alpha}-m_{V}$) for
the strongest H$\alpha$ emitters with model predictions. We then
applied the empirically determined zero-point offset to the entire
dataset, and found that for clusters younger than 10 Myr, with
measurable H$\alpha$ emission, age estimates were in good agreement
with those derived by Whitmore \& Zhang (2002). (The measured
H$\alpha$ fluxes for the clusters were converted to the {\sc vegamag}
system using the prescription given in the WFPC2 Data Handbook.)

We note that for those clusters without measurable H$\alpha$ EWs, we
applied the simplified method based on the broad-band luminosities
only, but using otherwise the same procedure.

\subsection{Method 4: Three-dimensional SED analysis (``3DEF'')}
\label{method4.sec}

The next step up in complexity of the fitting algorithms used involves
the fitting of the observed cluster SEDs to the Starburst99 and BC00
models using a three-dimensional (3D) maximum likelihood method, 3DEF,
with respect to a pre-computed grid of SSP models. This procedure was
described in detail by Bik et al. (2003), based on their analysis of
archival {\sl HST} observations of the central star clusters in M51,
and applied successfully to the intermediate-age star cluster system
in M82 B by de Grijs et al. (2003b) and to the extended cluster sample
in M51 by Bastian et al. (2004). The initial cluster mass $M_i$, age
and extinction E$(B-V)$ were adopted as free parameters. For those
clusters with upper limits in one or more filters, but still leaving
us with a minimum of three reliable photometric measurements, we use a
two-dimensional maximum likelihood fit (``2DEF''), using the
extinction probability distribution for E$(B-V)$. This distribution
was derived for the clusters with well-defined SEDs over the full
wavelength range (see Bik et al.  2003 for a full overview of this
procedure). The derivation of the most representative set of models
for a given cluster is done via a least-squares ($\chi^2$)
minimisation technique, in which the observed cluster SED is compared
to the full grid of SSP models. In the application of the 3DEF-method
Bik et al. (2003) and Bastian et al. (2004) assumed an uncertainty of
0.05 mag in the magnitudes of the cluster models (0.1 mag in the UV
filters).

\subsection{Method 5: Multidimensional SED analysis (``AnalySED'')}
\label{method5.sec}

Finally, we have developed a sophisticated SED analysis tool that can
be applied to photometric measurements in a given number $N (N \ge 4)$
of broad-band passbands (see de Grijs et al. 2003c, Anders et
al. 2004b). We apply a 3D $\chi^2$ minimisation to the SEDs of our
star clusters with respect to the {\sc galev} SSP models, to obtain
the most likely combination of age {\it t}, metallicity {\it Z} and
internal extinction E$(B-V)$ for each object (see Anders et al. 2004b;
Galactic foreground extinction is taken from Schlegel, Finkbeiner \&
Davis 1998).

In order to obtain useful results for all of our three free
parameters, i.e., age, metallicity and extinction\footnote{Strictly
speaking, the cluster mass is also a free parameter. Our model SEDs
are calculated for SSPs with initial masses of $1.6 \times 10^9
M_\odot$; to obtain the actual cluster mass, we scale the model SED to
match the observed cluster SED using a single scale factor. This scale
factor is then converted into a cluster mass.}, we need a minimum SED
coverage of four passbands.

Each of the models is assigned a probability, determined by a
likelihood estimator of the form $p \sim \exp(-\chi^2)$, where
\begin{equation}
\chi^2 \bigl(t,Z,{\rm E}(B-V),m_{\rm cl}\bigr)=\sum_{\rm models}
\frac{(m_{\rm obs}-m_{\rm model})^2}{\sigma^2_{\rm obs}}.
\end{equation}

Clusters with unusually large ``best'' $\chi^2$ are rejected, since
this is an indication of calibration errors, features not included in
the models (such as Wolf-Rayet star dominated spectra, objects younger
than 4 Myr, etc.) or problems due to the limited parameter
resolutions. We include an additional 0.1 mag per passband for ``model
uncertainties'' (0.2 mag for UV filters; see Section
\ref{models.sec}).

Subsequently, the model with the highest probability is chosen as the
``best-fitting model''. Models with decreasing probabilities are
summed up until reaching 68.26 per cent total probability (i.e., the 1
$\sigma$ confidence interval) to estimate the uncertainties on the
best-fitting model parameters. For each of these best-fitting models
the product of the relative uncertainties (${\rm \frac{age^+}{age^-}
\times \frac{mass^+}{mass^-} \times \frac{Z^+}{Z^-}}$) was calculated
(the superscripts indicate the upper ($^+$) and the lower limits
($^-$), respectively). The relative uncertainty of the extinction was
not taken into account, since the lower extinction limit is often
zero. For each cluster, the data set with the lowest value of this
product was adopted as the most representative set of parameters. In
cases where the analysis converged to a single model, a generic
uncertainty of 30 per cent was assumed for all parameters in linear
space, corresponding to an uncertainty of $^{+0.1}_{-0.15}$ dex in
logarithmic parameter space. See also de Grijs et al. (2003a,b,c) and
Anders et al. (2004a) for applications of this algorithm to NGC 3310
and NGC 6745, and NGC 1569, respectively, and Anders et al. (2004b)
for a theoretical analysis of its reliability.

We caution that the multi-passband combinations must not be biased to
contain mainly short wavelength, nor mainly long-wavelength
filters. Coverage of the entire optical wavelength range, if possible
with the addition of UV {\it and} NIR data, is most preferable (de
Grijs et al. 2003c, Anders et al. 2004b).

Finally, we emphasize once again that we will use this AnalySED method
as the basis for our comparisons among the different approaches
employed in this paper. This decision is based on the fact that the
method was validated and tested extensively, both empirically (de
Grijs et al. 2003a,c) and theoretically (Anders et al. 2004b), so that
we understand the systematic uncertainties inherent to this approach
in depth. In the following section, we will summarise the results from
our extensive validation of the AnalySED method, in order to justify
its use as our benchmark approach for comparison with the other
methods descirbed in the previous sections in the remainder of this
paper.

\section{Establishing our benchmark approach with artificial data}
\label{artdata.sec}

\begin{figure}
\includegraphics[width=8.6cm,angle=180]{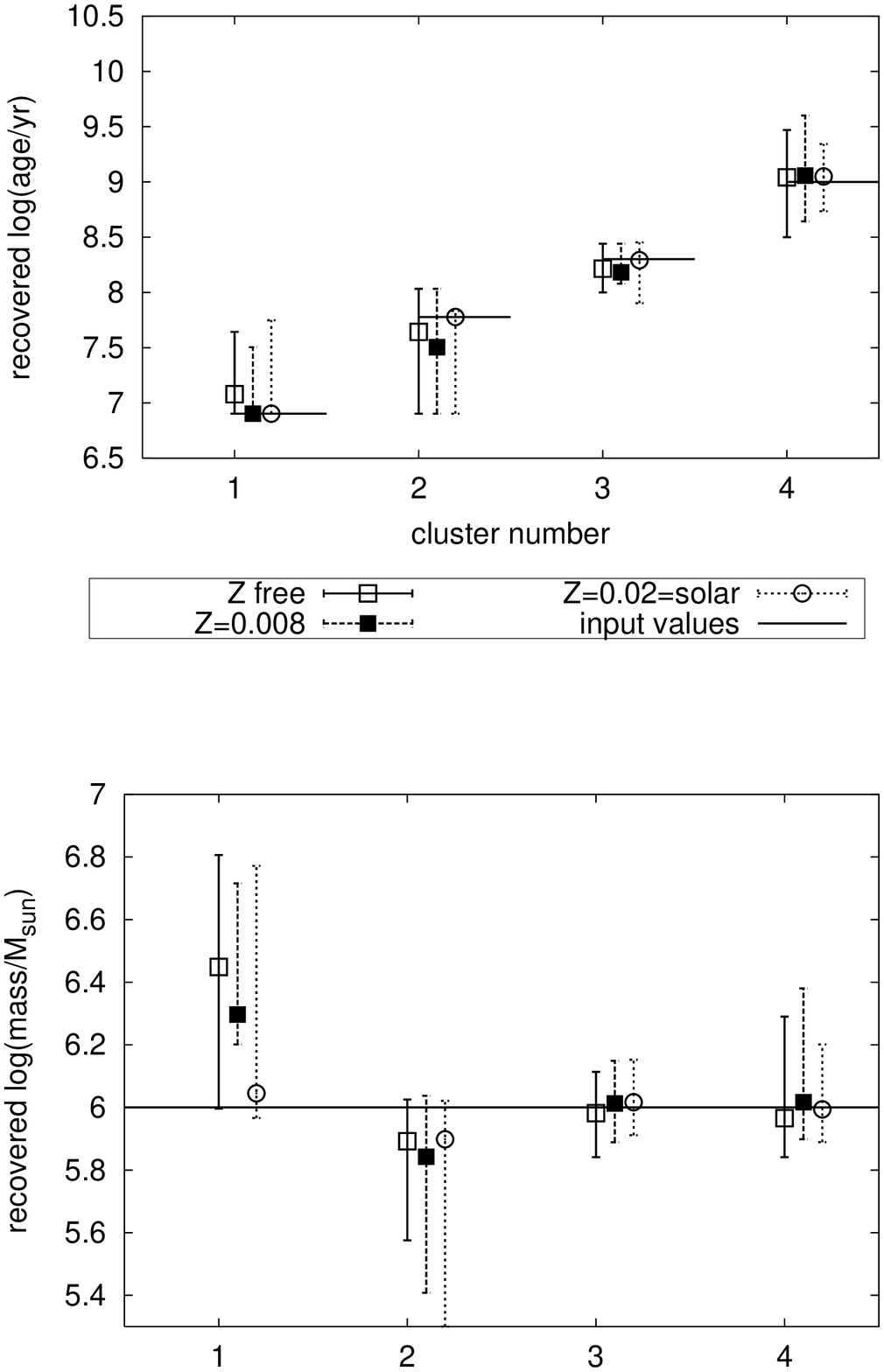}
\vspace{1cm}
\caption[]{Accuracy of the retrieval of the ages and masses of
artificial star clusters based on a wavelength coverage as available
for our NGC 3310 cluster sample. The artifical clusters are
characterised by E$(B-V) = 0.1$ mag, $Z = Z_\odot$, and ages of 8, 60,
and 200 Myr, and 1 Gyr (objects 1--4). The different symbols represent
the retrieved values based on a variety of a priori assumptions on the
clusters' metallicities.}
\label{art3310.fig}
\end{figure}

\begin{figure}
\includegraphics[width=8.6cm,angle=180]{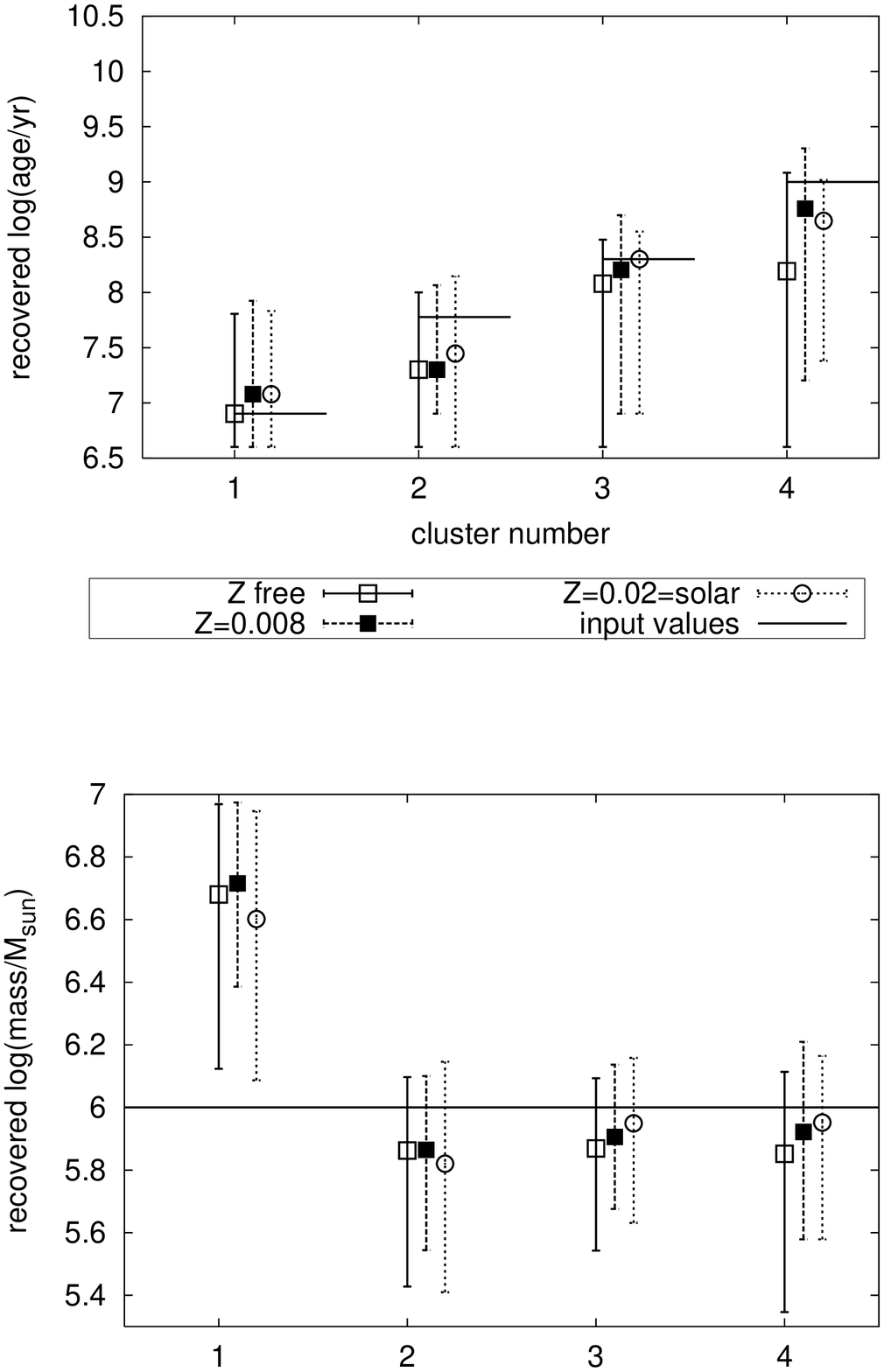}
\vspace{1cm}
\caption[]{Accuracy of the retrieval of the ages and masses of
artifical star clusters based on a wavelength coverage as available
for our Antennae cluster sample; technical details are as in
Fig. \ref{art3310.fig}.}
\label{art4038.fig}
\end{figure}

In Anders et al. (2004b) we presented a detailed study of the
reliability and limitations of our AnalySED approach. We computed a
large grid of broad-band {\sl HST}-based star cluster SEDs on the
basis of our {\sc galev} models for SSPs, including all relevant
up-to-date input physics for stellar ages $\ge 4$ Myr. We constructed
numerous artificial cluster SEDs, and varied each of the input
parameters (specifically, age, metallicity, and internal extinction;
see Section \ref{method5.sec}) in turn to assess their effects on the
robustness of our parameter recovery. For each clean model artificial
cluster SED we calculated 10,000 additional clusters, with errors
distributed around the input magnitudes in a Gaussian fashion.

By analysing artificial clusters, using a variety of input parameters
with our AnalySED approach, we found in general good agreement between
the recovered and the input parameters for ages $\lesssim 10^9$ yr,
i.e., exactly the age range of interest for the clusters in NGC 3310
and the Antennae galaxies analysed in this paper.

We considered several a priori restrictions of the full parameter
space, both to the (correct) input values and to some commonly assumed
values. We easily recover all remaining input values correctly if one
of them is restricted a priori to its correct input value; this also
provides a sanity check for the reliability of our code. We conclude
that the age-metallicity degeneracy is responsible for some
misinterpretations of clusters younger than $\sim 200$ Myr. If we
restrict one or more of our input parameters a priori to incorrect
values (such as by using, e.g., only solar metallicity, as often done
in the literature), large uncertainties result in the remaining
parameters.

In order to provide a robust theoretical benchmark for the
observational study of systematic uncertainties presented in the
remainder of this paper, here we re-validate our AnalySED approach
using the exact filter combinations available for our NGC 3310 and
Antennae cluster samples (Tables \ref{ngc3310.tab} and
\ref{antennae.tab}), i.e., by computing a large grid of broad-band
star cluster SEDs in a similar fashion as done in Anders et
al. (2004b) -- although using only 1,000 additional artificial
clusters to quantify our model uncertainties; the difference between
this approach and the 10,000 additional artificial clusters used in
Anders et al. (2004b) is negligible, however. Our artifical cluster
SEDs were computed for ages of 8, 60, and 200 Myr and 1 Gyr (the age
range covered by our sample clusters; see Sections \ref{n3310cov.sec}
and \ref{n4038cov.sec}), and for a fixed (internal) extinction of
E$(B-V) = 0.10$ mag and solar metallicity. The extinction value
adopted roughly corresponds to the mean extinction derived for the
individual clusters (Sections \ref{n3310cov.sec} and
\ref{n4038cov.sec}); extinction variations among the sample clusters
are small, and their effects (for the derived range of extinction
values) on the age and mass estimates are negligible (de Grijs et
al. 2003a, Anders et al. 2004b). The adopted solar metallicity also
corresponds roughly to the mean metallicity derived for the sample
clusters, although it may not be correct in individual cases. In order
to quantify the effects of metallicity variations, we attempted to
retrieve the ages and masses of our artificial clusters by assuming
both the correct (solar) and incorrect ($Z = 0.008 = 0.4 Z_\odot$)
metallicities as a priori restrictions. In addition, we retrieved the
ages and masses of the artificial clusters without any restriction to
the resulting metallicity (``$Z$ free''). The latter provides a
quantitative indication of the importance of the age-metallicity
degeneracy.

The results of this re-validation are shown in Figs. \ref{art3310.fig}
and \ref{art4038.fig} for filter coverage as for NGC 3310 and the
Antennae galaxies, respectively. It is immediately clear that the
accuracy of the parameter retrieval is significantly better for the
NGC 3310 clusters than for those in the Antennae galaxies, which
simply reflects the available filter sets (cf. de Grijs et
al. 2003a,c, Anders et al. 2004b). Nevertheless, in all cases the {\it
ages} are retrieved well within the modelling uncertainties, for any
assumption on the clusters metallicity. For the coverage corresponding
to the NGC 3310 clusters, the difference between input and retrieved
ages is $\Delta \log({\rm Age/yr}) \lesssim 0.3$, and in the majority
of cases $\Delta \log({\rm Age/yr}) \lesssim 0.15$. Except for the 60
Myr-old cluster, $\Delta \log({\rm Age/yr}) \lesssim 0.3$ also for
those clusters covered by the same passbands as the Antennae clusters
(although the age-metallicity degeneracy is somewhat more important
for the 1 Gyr-old cluster in this case). The corresponding uncertainty
in the retrieved age of the 60 Myr-old artificial cluster is about
twice as large as for the other clusters; its cause is unclear, since
the retrieved extinction and metallicity values for this object are
not significantly more uncertain than for the other objects.

A similar behaviour, i.e., with slightly larger uncertainties for the
Antennae-equivalent wavelength coverage compared to the coverage of
the NGC 3310 sample, is seen for the retrieved {\it masses} of the
artificial clusters, although to a lesser extent. The mass
uncertainties for all clusters older than 60 Myr are $\Delta \log(
M_{\rm cl}/M_\odot ) \lesssim 0.10-0.15$. For the youngest, 8 Myr-old
object, we are only able to retrieve the masses to within several 0.1
dex in mass (somewhat more accurately for the NGC 3310-equivalent
wavelength coverage, particularly if the adopted metallicity is close
to the actual value); this is most likely caused by the uncertainties
inherent to our present knowledge of stellar evolution in this age
range (such as, e.g., the importance of the RSG phase), and the
relatively coarse age resolution compared to the rapidity of changes
in stellar evolution around 6--12 Myr.

Thus we have shown, based on well-understood artificial data, that we
understand {\it quantitatively} the uncertainties inherent to using
our AnalySED approach for age and mass determinations of star clusters
based on broad-band imaging. We will be using this approach as our
benchmark for comparing our results to those obtained using
alternative methods commonly in use in the community. 

In Section \ref{models.sec} we concluded that the differences among
the various prescriptions used for the input physics in modern sets of
SSP models are very small indeed and apparently not biased
systematically. As a consequence of the analysis performed in this
section, we conclude then that any differences in the individual (as
well as in the mean) cluster ages and masses that we will find in
subsequent sections (over and above the modelling uncertainties
quantified here) are most likely caused by intrinsic differences among
the various {\it methods}.

\section{Comparison of the relative age and mass distributions}
\label{comparison.sec}

\subsection{Extensive wavelength coverage: NGC 3310}
\label{n3310cov.sec}

To start our comparison of methods, we will focus on the extensive
wavelength coverage of the NGC 3310 star cluster system. With coverage
from the F300W {\sl HST} mid-UV passband to the NIR F205W passband,
the resulting broad-band SEDs were shown to have sufficient leverage
to distinguish metallicity, extinction, and stellar population (age)
effects (de Grijs et al. 2003c).

While a wavelength coverage as extensive as possible is preferred, the
use of {\sl HST}-flight system magnitudes (cf. Section \ref{data.sec})
limits the application of the NGC 3310 comparison to the use of the
{\sc galev} SSP models (see Section \ref{galev.sec}), which we folded
through the {\sl HST}/WFPC2 filter curves ourselves (e.g., de Grijs et
al. 2003c, Bastian et al. 2004).

We emphasize, however, that we prefer to use the original {\sl
HST}-flight filter system, rather than conversions to ``standard''
systems; in Section \ref{filters.sec} we will discuss the systematic
effects unavoidably introduced when converting {\sl HST}-flight
system magnitudes to the ``standard'' Johnson-Cousins system.

Figure \ref{n3310ages.fig} shows the resulting age distributions for
the 17 NGC 3310 star clusters used for this exercise, obtained using a
variety of approaches. In Fig. \ref{n3310ages.fig}a, we display the
relative age distribution of the NGC 3310 clusters based on the full
multi-dimensional SED analysis (Section \ref{method5.sec}), in which
we left all of the cluster ages, masses, metallicities and extinction
values as free parameters. The vertical dotted line at log(Age/yr) =
6.6 denotes the lower age limit of the {\sc galev} SSP models; the
vertical error bars indicate the Poissonian uncertainties.

\begin{figure}
\psfig{figure=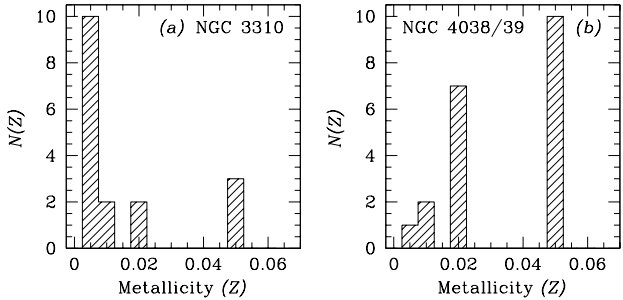,width=9cm}
\vspace{-4.3cm}
\caption[]{Metallicity estimates of the NGC 3310 and Antennae sample
clusters, based on the results obtained from the AnalySED
multi-dimensional approach. Note that these are first-order estimates
(yielding an internally consistent overall metallicity distribution;
see Anders et al. 2004b), and that the uncertainties in the individual
cluster metallicities are on the order of one step in our metallicity
grid, covering metallicities of $0.004 Z_\odot, 0.2 Z_\odot, 0.4
Z_\odot, Z_\odot (Z=0.02)$, and $2.5 Z_\odot$.}
\label{metallicities.fig}
\end{figure}

\begin{figure}
\psfig{figure=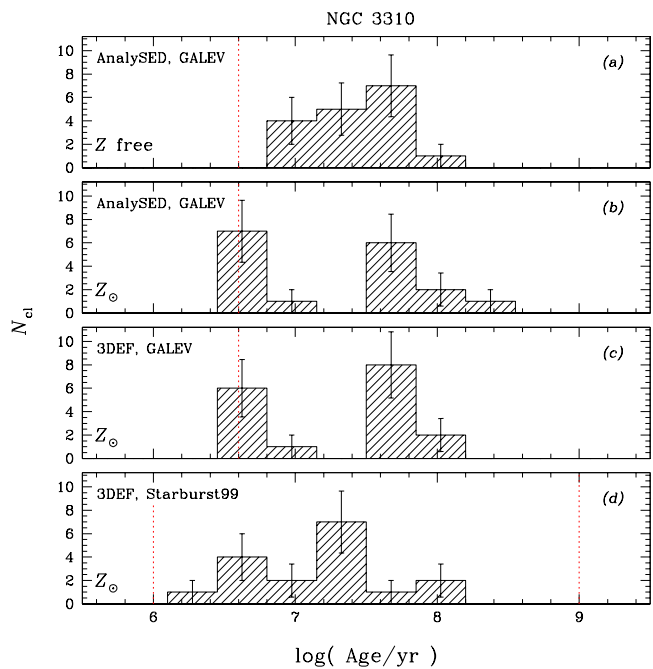,width=9cm}
\caption[]{Resulting age distributions of the NGC 3310 star clusters,
using a variety of methods and SSP models, as indicated in the
individual panels. The vertical dotted lines correspond to the fitting
boundaries of the models. Metallicities are indicated in the panels.}
\label{n3310ages.fig}
\end{figure}

Panel (b) provides a direct comparison of the effects of metallicity
variations. Here, as well as in the other panels in this figure, we
have adopted solar metallicity for the individual star clusters. While
this is not necessarily correct in general, restricting the
metallicity to the solar value allows a more robust comparison among
the various models and methods\footnote{This is because by treating
metallicity as a fit parameter we introduce the complexity of an
additional free parameter, which has the potential to render the
computational solution less stable and robust. In view of the small
effects associated with small differences in metallicity, for the
purpose of this exercise, we opt for the more robust approach to adopt
a single (solar) metallicity. We note that this reflects common
practice in the literature (but see de Grijs et al. 2003c).}. The
differences in the relative age distributions between
Figs. \ref{n3310ages.fig}a and b are therefore entirely and
exclusively due to the different assumptions on the clusters'
metallicities. They reflect the well-known effects of the
age-metallicity degeneracy (e.g., Ferreras \& Yi 2004; see de Grijs et
al. [2003a,c] for detailed studies of this effect in NGC 3310 and NGC
6745). Based on the multi-dimensional analysis presented in
Fig. \ref{n3310ages.fig}a, most (70 per cent) of the NGC 3310 clusters
in our current sample are characterised by metallicities $Z \lesssim
0.01$, with the remainder split evenly between solar metallicity ($Z =
0.02$) and $Z \sim 0.05$, see Fig. \ref{metallicities.fig}. Support
for these metallicity estimates is provided by the unusually low
(subsolar) metallicity found independently in star-forming regions
surrounding the nucleus of NGC 3310, while the nucleus itself appears
to have solar metallicity (e.g., Heckman \& Balick 1980, Puxley,
Hawarden \& Mountain 1990, Pastoriza et al.  1993).  Similarly, the
age-extinction degeneracy may contribute to some extent, although its
effect is likely less than that of the age-metallicity degeneracy
(cf. de Grijs et al. 2003c). The extinction in this sample of NGC 3310
clusters, E$(B-V)$, decreases with age -- although with a large
scatter -- from E$(B-V) \sim 0.35$--0.40 mag at $\sim 10^{6.5}$ yr, to
E$(B-V) < 0.1$ mag at $\sim 10^8$ yr.

Figure \ref{n3310ages.fig}c can be compared directly with
Fig. \ref{n3310ages.fig}b; the only difference between these two
panels is that we used the ``3DEF'' method (Section \ref{method4.sec})
instead of the AnalySED multi-dimensional approach. We used the {\sc
galev} SSP models in both cases. It is encouraging to see that the use
of either method results in very similar relative age
distributions. More quantitatively, the two peaks in the age
distribution are reproduced to a very high degree of confidence ($>
99$ per cent), although their relative amplitudes are subject to
small-number statistics; as a result, the straightforward application
of a Kolmogorov-Smirnov (KS) test\footnote{From a pure statistics
approach, the application of a KS test to our results is strictly
speaking invalid. While the relatively small number of data points is
in principle acceptable and sufficient for our purpose, systematic
effects related to the sample selection are not taken into account,
while they may in fact dominate. This thus renders the results not
very illustrative.} yields a probability that these data points were
drawn from statistically different distributions of $\sim 80$ per
cent, but with a very large uncertainty because of the small number of
data points used.

\begin{table}
\caption[ ]{\label{n3310.tab}Characteristics of the overall relative
mass distributions of our NGC 3310 cluster sample}
{\scriptsize
\begin{center}
\begin{tabular}{llccc}
\hline
\hline
\multicolumn{1}{c}{Method} & \multicolumn{1}{c}{SSPs} &
\multicolumn{1}{c}{$Z$} & 
\multicolumn{2}{c}{log(Mass/M$_\odot$)}\\
\cline{4-5}
& & & \multicolumn{1}{c}{Mean} & \multicolumn{1}{c}{$\sigma$} \\
\hline
%AnalySED & {\sc galev} & free      & 7.43 & 0.33 && 5.13 & 0.28 \\
%AnalySED & {\sc galev} & $Z_\odot$ & 7.28 & 0.65 && 4.92 & 0.29 \\
%3DEF     & {\sc galev} & $Z_\odot$ & 7.33 & 0.59 && 4.89 & 0.23 \\
%3DEF     & Starburst99 & $Z_\odot$ & 7.16 & 0.49 && 5.04 & 0.24 \\
AnalySED & {\sc galev} & free      & 5.13 & 0.28 \\
AnalySED & {\sc galev} & $Z_\odot$ & 4.92 & 0.29 \\
3DEF     & {\sc galev} & $Z_\odot$ & 4.89 & 0.23 \\
3DEF     & Starburst99 & $Z_\odot$ & 5.04 & 0.24 \\
\hline
\end{tabular}
\end{center}
}
\end{table}

Finally, in panel (d) we have replaced the {\sc galev} SSP models by
the Starburst99 models (Section \ref{sb99.sec}), which results in a
markedly different age distribution. This is most likely caused by two
effects, which are in essence the most significant differences between
these two sets of SSP models. The {\sc galev} SSP models include the
contributions of an extensive set of nebular emission lines and
gaseous continuum emission, which have been shown to be important in
the first few $\times 10^7$ yr of an SSP's evolution (Anders \&
Fritze-v. Alvensleben 2003); the Starburst99 set of SSP models does
include nebular continuum emission, but only in a simplified
fashion. The other main difference between both sets of SSP models is
related to their treatment of the RSG phase, which is of significant
importance around $10^7$ yr: it is clear from panels (b) and (c) that
using the {\sc galev} models leaves a gap in the clusters' age
distribution around the time that the RSG phase is expected to be
important. This is caused by a combination of both the age-metallicity
degeneracy and the sparser age resolution of the {\sc galev} models
compared to that of the Starburst99 SSPs. Lamers et al. (2001) have
shown that the Geneva models of fully convective stars are not cool or
red enough compared to the observations, particularly at lower
metallicities (see also Massey \& Olsen 2003). This may have
significant consequences for techniques which allow the metallicity to
be a free parameter, obviously depending on the age range of the
respective clusters. Whitmore \& Zhang (2002) have shown that cluster
models calculated with the Padova tracks fit the observations better
than those calculated with the Geneva tracks.

\begin{figure}
\psfig{figure=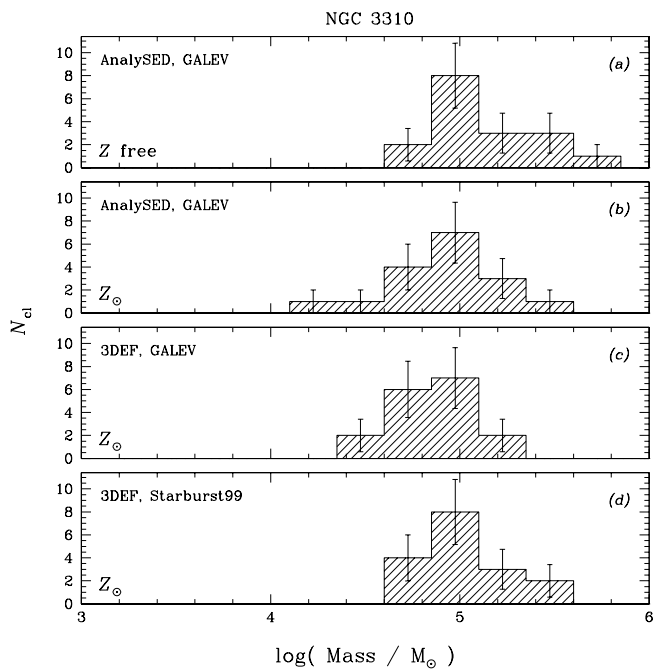,width=9cm}
\caption[]{Resulting mass distributions of the NGC 3310 star clusters,
using a variety of methods and SSP models, as indicated in the
individual panels. Metallicities are also indicated in the panels.}
\label{n3310masses.fig}
\end{figure}

The relative mass distributions, shown in Fig. \ref{n3310masses.fig},
are much more similar to each other than the corresponding age
distributions when we compare the various methods and SSP models
used. The cluster masses have all been corrected to the stellar IMF
used by the {\sc galev} SSPs (see Section \ref{models.sec}). The
effect of the age-metallicity degeneracy is seen to some extent
between panels (a) and (b,c,d): as we established above, this
degeneracy causes the cluster ages to be underestimated if the
metallicity is overestimated (as is likely the case if we assume solar
metallicity for the NGC 3310 clusters), which in turn causes the
cluster masses to be underestimated. However, this effect is minor in
our NGC 3310 cluster sample. In spite of the inherent problems of
applying KS tests to astrophysical data such as presented here, the
results of such tests are illustrative in a comparative fashion: the
difference in the adopted metallicity reduces the probability of
distributions (a) and (b) to have been drawn from the same population
to only 19.0 per cent. The agreement between distributions (b), and
(c) and (d), are more satisfactory, with probabilities of these having
been drawn from the same population of 93.0 and 67.3 per cent,
respectively. In addition to the KS statistics, we can also compare
the overall statistics of the distributions in Fig.
\ref{n3310masses.fig}, as shown in Table \ref{n3310.tab}, which shows
that we can reproduce the mean (``peak'') and spread of the
distributions consistently and well within the uncertainties
(represented by the $\sigma$ values), even where we adopted different
metallicity distributions; if we simply compare the peak values of the
mass distributions obtained from the various methods, we find a spread
among these values of $\sigma_M \equiv \Delta \langle \log( M_{\rm cl}
/ {\rm M}_\odot ) \rangle \le 0.06$ (where we have only used the peak
values obtained assuming similar boundary conditions, i.e., for
$Z_\odot$; the peak values were obtained from Gaussian fits to the
distributions of the individual cluster masses). If we had simply
taken the mean of the age distributions and done the same comparison,
the resulting spread would have been $\sigma_t \equiv \Delta \langle
\log( {\rm Age / yr} ) \rangle \le 0.15$ (although we note that this
result is unphysical, in view of the significantly non-Gaussian
distributions, so that a single ``mean'' value does not convey much
useful information). Thus, we conclude that the {\it peaks} in the
relative mass distributions can be derived much more consistently than
those in the relative age distributions. This is caused by the way in
which the masses are determined (by scaling up either the entire
observed SED or the $V$-band flux to the models), and the fairly
narrow age range covered by the NGC 3310 clusters (ensuring a
relatively small range of cluster mass-to-light ratios).

\subsection{Restricted wavelength coverage, including H$\alpha$ observations: 
NGC 4038/9}
\label{n4038cov.sec}

While ideally one would like to have the most extensive wavelength
coverage possible, realistically one cannot expect to obtain more {\sl
HST} coverage than by, say, four passbands for any given cluster
system. Under this assumption, we have shown -- both empirically (de
Grijs et al. 2003a,c) and theoretically (Anders et al. 2004b) -- that
the optical $UBVI$ passbands (or their equivalents in the {\sl HST}
flight system) provide the most suitable passband combination to use
as the basis for our broad-band SED analysis. Additional NIR
observations would add significantly more leverage, but in practice
such observations need to be obtained using different detectors, and
are thus more difficult to obtain. We focus therefore on data sets
that can be obtained with minimal observing time, while maximising the
scientific output.

In this section, we will explore the differences between the various
methods, using $UBVI$ coverage of 20 star clusters in the Antennae
interacting galaxies (NGC 4038/9), selected to span a large age range
in the original analysis where their properties were first published
(see Whitmore et al. 1999, and references therein). The ``standard''
$UBVI$ cluster magnitudes were obtained by converting the {\sl
HST}-flight system magnitudes using the Holtzman et al. (1995)
conversion equations. For a subset of these clusters we have also
obtained H$\alpha$ EWs, which can -- in principle -- be used to
constrain their ages more accurately and robustly, although strong and
patchy background H$\alpha$ fluxes (as observed in the Antennae
galaxies) can render the actual contributions to the H$\alpha$ fluxes
by the clusters themselves very uncertain.

\begin{figure}
\psfig{figure=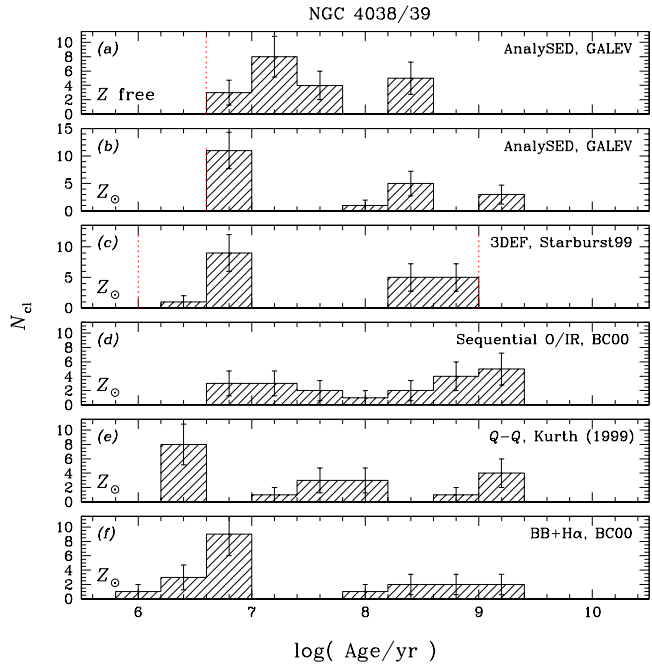,width=9cm}
\caption[]{Resulting age distributions of the Antennae star clusters,
using a variety of methods and SSP models, as indicated in the
individual panels. The vertical dotted lines correspond to the
boundaries of the models. Metallicities are indicated in the panels.}
\label{n4038ages.fig}
\end{figure}

Figure \ref{n4038ages.fig} shows the relative age distributions
resulting from the application of the various methods described in
Section \ref{methods.sec}. The top three panels of
Fig. \ref{n4038ages.fig}, (a), (b), and (c), show similar trends as
pointed out for the same method+models combinations used for the NGC
3310 clusters in the previous section. The effects of the
age-metallicity degeneracy are somewhat less pronounced in this case,
since our multi-dimensional AnalySED analysis (Section
\ref{galev.sec}) indicates that the cluster metallicities in the
Antennae galaxies are roughly equally split between solar metallicity
and $Z \sim 2.5 Z_\odot$ (with only a few clusters characterised by
subsolar metallicities, $Z \lesssim 0.01$). These metallicity
estimates are supported by high-resolution spectroscopy obtained by
Mengel et al. (2002). Thus, by assuming solar metallicity for all
Antennae clusters, we will have overestimated the ages for those
clusters with supersolar metallicity, and underestimated the ages of
the few subsolar metallicity clusters. This is reflected by the
different age distributions between panels (a) vs. (b) and (c). Again,
application of these methods leaves a clear gap in the age
distributions around the age where the RSGs become apparent.

The multi-dimensional SED analysis further shows a weak correlation
between E$(B-V)$ and cluster age (although with a large scatter), from
E$(B-V) \lesssim 1$ mag at $\sim 10^{6.5}$ yr to E$(B-V) \lesssim 0.4$
mag at $\sim 10^{9.5}$ yr.

In Fig. \ref{n4038ages.fig}d, we display the results from the
``Sequential O/IR'' method (Section \ref{method1.sec}).  Based on the
availability of $BVI$ photometry, this method also allows us to
determine the extinction toward the sample clusters
independently. The weak trend found by our multi-dimensional SED
approach is also found using this method, although the individual
extinction values are generally slightly smaller, E$(B-V) \lesssim
0.2$ mag.

Yi et al. (2004) have shown that the $(U-B)$ versus $(B-V)$ two-colour
diagram can be used to break the age-metallicity degeneracy for
metal-poor populations. However, since our sample clusters are metal
rich, we cannot apply this technique here. All of their ageing
trajectories (based on the BC00 SSP models), whatever their
metallicity, are at the same locus of the diagram, at least for
stellar populations older than 100 Myr.

While the double-peaked age distribution obtained in panels (a)--(c)
is to some extent reproduced, the two-step process of the Sequential
O/IR method results in a more evenly spread age distribution. This is
partially due to the fact that some of the sample clusters are
apparent outliers in the diagnostic diagrams, and as a consequence
their ages are not well constrained.

Finally, Figs. \ref{n4038ages.fig}e and f show the age distributions
resulting from using the reddening-free $Q$-parameter analysis and the
broad-band+H$\alpha$ method (Sections \ref{method2.sec} and
\ref{method3.sec}), respectively. These distributions are, in very
broad terms, consistent with those obtained using the other methods
discussed before, in the sense that they show multiple peaks at
roughly similar ages (although they do not match in detail). The most
deviant distribution is in fact that resulting from the ``Sequential
O/IR'' method, which is based on a smaller number of photometric data
points per cluster than the other methods.

\begin{figure}
\psfig{figure=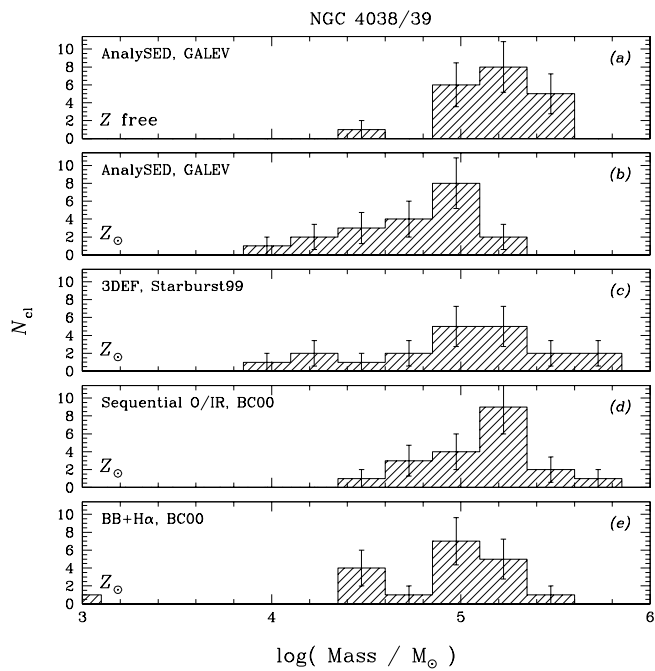,width=9cm}
\caption[]{Resulting mass distributions of the Antennae star clusters,
using a variety of methods and SSP models, as indicated in the
individual panels. Metallicities are also indicated in the panels.}
\label{n4038masses.fig}
\end{figure}

Figure \ref{n4038masses.fig} shows the corresponding mass distribution
for our sample of Antennae clusters, based on a variety of
methods. Because of the small-number statistics and, in particular,
the different mass range covered by all of the mass distributions in
Fig. \ref{n4038masses.fig} the simple KS statistic indicates that all
distributions are different from our base line distributions,
Figs. \ref{n4038masses.fig}a and b, at the $>98$ per cent level (even
if we restrict ourselves to the largest possible mass range in common
between any two sets of mass determinations, and common metallicity
assumptions). However, as for NGC 3310, the overall characteristics of
the mass distributions are fairly consistently reproduced, in
particular the mean mass (see also Table \ref{n4038.tab}); similarly
as for NGC 3310, we find a spread in the mean mass among the various
approaches of $\sigma_M \equiv \Delta \langle \log( M_{\rm cl} / {\rm
M}_\odot ) \rangle \le 0.14$ (once again, for the fits done assuming
$Z_\odot$ only). The equivalent spread for the (unphysical) mean in
the age distributions would be $\sigma_t \equiv \Delta \langle \log(
{\rm Age / yr} ) \rangle \le 0.35$. It appears, therefore, that --
once more -- the {\it peaks} in the relative mass distributions (as
opposed to the detailed shapes of the distributions\footnote{For a
comparison of the detailed shape of the underlying distribution, one
would need a statistically much larger (and unbiased) sample of
clusters than studied here. It should be noted that here, we selected
clusters biased in such a way that they would cover as extensive a
range in ages and masses as possible for these galaxies, based on
preliminary analysis (see Section \ref{data.sec}). For smaller samples
such as ours, the uncertainties in the {\it individual} age and mass
estimates (which are on the order of the histogram bin sizes; cf. de
Grijs et al. 2003a,c, Anders et al. 2004b) start to affect the
resulting distribution non-negligibly and in ways that cannot easily
be quantified.}) can be obtained with a much higher degree of
confidence than those of the relative age distributions.

The overall characteristics of the mass distributions for the current
sample of Antennae clusters are summarised in Table \ref{n4038.tab},
which were obtained using the same fitting algorithms as for the fits
to the NGC 3310 mass distributions. We note that, due to insufficient
information in the Kurth et al. (1999) SSP models at the highest time
resolution (used in this paper), we were unable to determine the
cluster masses for the $Q-Q$ method.

\begin{table}
\caption[ ]{\label{n4038.tab}Characteristics of the overall relative
mass distributions of our Antennae cluster sample}
{\scriptsize
\begin{center}
\begin{tabular}{llccc}
\hline
\hline
\multicolumn{1}{c}{Method} & \multicolumn{1}{c}{SSPs} &
\multicolumn{1}{c}{$Z$} & 
\multicolumn{2}{c}{log(Mass/M$_\odot$)}\\
\cline{4-5}
& & & \multicolumn{1}{c}{Mean} & \multicolumn{1}{c}{$\sigma$} \\
\hline
%AnalySED    & {\sc galev}    & free      & 7.50 & 0.55 && 5.20 & 0.26 \\
%AnalySED    & {\sc galev}    & $Z_\odot$ & 7.58 & 1.07 && 4.72 & 0.31 \\
%3DEF        & Starburst99    & $Z_\odot$ & 7.74 & 1.02 && 4.97 & 0.47 \\
%Seq. O/IR   & BC00           & $Z_\odot$ & 8.11 & 0.94 && 5.12 & 0.32 \\
%{\it Q--Q}  & Kurth (1999)   & $Z_\odot$ & 7.63 & 1.10 \\
%BB+H$\alpha$ & BC00          & $Z_\odot$ & 7.36 & 1.04 && 4.91 & 0.32 \\
AnalySED    & {\sc galev}    & free      & 5.20 & 0.26 \\
AnalySED    & {\sc galev}    & $Z_\odot$ & 4.72 & 0.31 \\
3DEF        & Starburst99    & $Z_\odot$ & 4.97 & 0.47 \\
Seq. O/IR   & BC00           & $Z_\odot$ & 5.12 & 0.32 \\
{\it Q--Q}  & Kurth (1999)   & $Z_\odot$ \\
BB+H$\alpha$ & BC00          & $Z_\odot$ & 4.91 & 0.32 \\
\hline
\end{tabular}
\end{center}
}
\end{table}

\section{One-to-one comparisons of mass and age estimates}
\label{detailed.sec}

Having established that the peak (or most frequent value) of the
relative age and, in particular, mass distributions of a given star
cluster system can be retrieved relatively robustly using broad-band
SEDs, we will now explore to what extent this applies to the
individual cluster properties themselves. It is clear from the outset
that where multiple estimates for the cluster extinction values, and
to a lesser extent also for their metallicities, exist, these
estimates vary significantly. However, despite this being the case
from an observational point of view, the effects on the SED of varying
the extinction and metallicity, if their uncertainties are not more
than a few tenths in magnitude or one step in metallicity,
respectively, are small, and can thus still result in reasonably
secure age and mass estimates for all but the youngest clusters
(cf. Anders et al. 2004b). In other words, the age and mass estimates
are relatively robust to variations in the extinction and metallicity
estimates. Therefore, we will only discuss the one-to-one comparisons
of the ages and masses of the individual clusters in either of our
samples. We also note that the youngest SEDs ($\lesssim 10^7$ yr) can
be affected quite significantly (and not in any systematic fashion) by
even small changes in extinction and/or metallicity, however (see
Anders et al. 2004b).

\begin{figure}
\psfig{figure=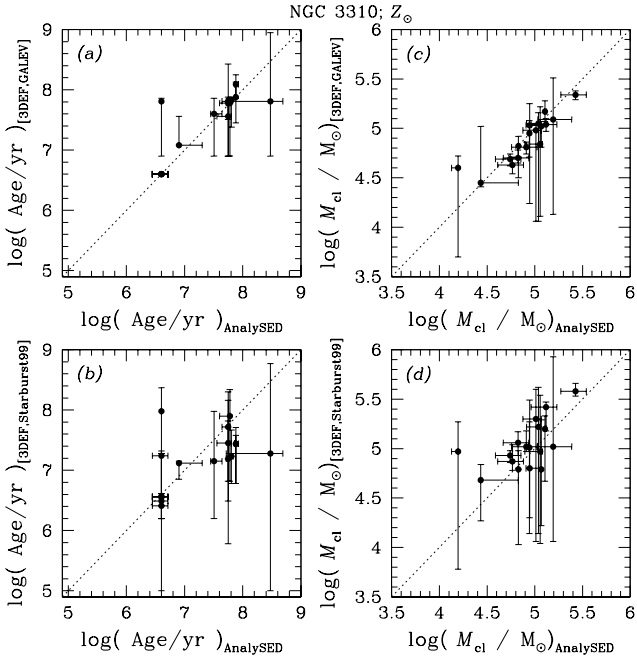,width=9cm}
\caption[]{One-to-one comparisons between the age (left) and mass
(right) estimates of the NGC 3310 clusters, for solar metallicity,
obtained from the various methods and SSP models, using our AnalySED
approach as the basis for comparison. The dotted lines are the loci of
equality.}
\label{cf3310.fig}
\end{figure}

Figure \ref{cf3310.fig} shows the one-to-one comparisons of the NGC
3310 cluster ages and masses, adopting solar metallicity, and using
the AnalySED approach as our basis for the comparison. It is
immediately clear that the methods using the same set of SSPs,
Figs. \ref{cf3310.fig}a and c for the ages and masses, respectively,
result in highly reproducible age and mass estimates (well within the
1-$\sigma$ error bars). With few exceptions, the individual ages and
masses match very well within the model uncertainties.

In the most extreme case considered for the NGC 3310 star cluster
system, namely by using a different method (AnalySED vs. 3DEF) {\it
and} a different set of SSP models ({\sc galev} vs. Starburst99; shown
in Figs. \ref{cf3310.fig}b and d), the individual age and mass
estimates still match up well within the model uncertainties, although
with a slightly larger scatter. The ``3DEF,Starburst99'' approach
results in slightly higher masses than the AnalySED approach, although
this is only a $\sim 1 \sigma$ deviation.

\begin{figure}
\psfig{figure=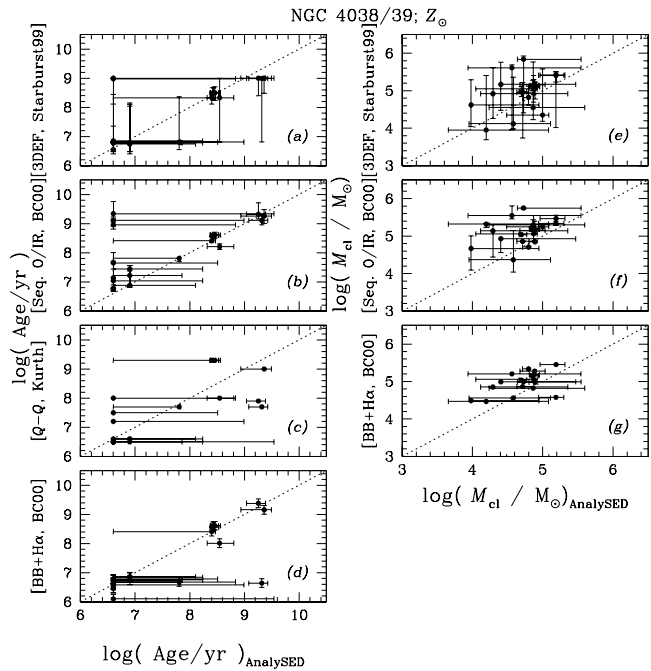,width=9cm}
\caption[]{One-to-one comparisons between the age (left) and mass
(right) estimates of the Antennae clusters, for solar metallicity,
obtained from the various methods and SSP models, using our AnalySED
approach as the basis for comparison. The dotted lines are the loci of
equality.}
\label{cf4038.fig}
\end{figure}

Figure \ref{cf4038.fig} shows a similar set of comparisons for the
Antennae clusters, now using the entire range of methods and models at
our disposal. At first sight, we notice three characteristics when
comparing these panels to the panels of Fig. \ref{cf3310.fig}. First,
the uncertainties in the ages and masses estimated by the AnalySED
approach (see Anders et al. 2004b for a full description and
justification) are significantly larger than those resulting from most
other methods used, except for the 3DEF approach. Secondly, the
uncertainties are much greater than those for the individual age and
mass estimates obtained for the NGC 3310 clusters. This reflects the
difference in wavelength coverage of the broad-band SEDs for both sets
of clusters: with the smaller wavelength coverage of the Antennae
clusters, their best-fitting ages and masses are not as well
constrained as for the more extensively covered NGC 3310 clusters. It
therefore appears that the uncertainties estimated using the
Sequential O/IR, the {\it Q--Q} and the BB+H$\alpha$ methods are too
small in view of the ill-defined broad-band SEDs over the fairly small
wavelength range covered. We should point out, however, that the
uncertainties resulting from the BB+H$\alpha$ method (i.e., for
clusters with measured H$\alpha$ EWs) are likely to be smaller than
those from the broad-band-only methods, since this method uses
H$\alpha$ EWs to further constrain the cluster ages. This method is
therefore most accurate in a very narrow age range (i.e., more
accurate than methods based on broad-band photometry alone), around
$\sim 10$ Myr. Finally, it appears that the mass estimates of the
individual clusters are better matched than their age estimates. This,
in turn, provides the more robust {\it relative} mass distributions
discussed in the previous Sections.

Thus, we believe that the main differences among the resulting age and
mass estimates for individual clusters are caused by the difference in
the wavelength range covered, but -- as we will show in Section
\ref{filters.sec} -- the uncertainties introduced by transforming the
{\sl HST}-flight system magnitudes to ``standard'' ground-based {\it
UBVI} photometry may affect the robustness of the results to a
similar, if not greater, degree.  While the individual cluster ages
and masses based on the ``ideal'' $UBVI$ (or equivalent) coverage
advocated in de Grijs et al. (2003c), and Anders et al. (2004b) may be
subject to significant uncertainties, the relative age and mass
distributions are much more consistently established, so that
statistical analyses of large-scale cluster systems based on
multi-passband broad-band imaging do have the potential to provide
robust scientific insights in their host galaxies' formation,
evolution, and star-formation histories.

Finally, in Table \ref{ha-ages.tab} we compare the age estimates for
the individual clusters in our Antennae cluster sample with H$\alpha$
EW $> 1.0$\AA, for the full set of SSP analysis methods used. We have
indicated the clearly discrepant age estimates in italic font. The
final two columns of this table include the mean age and its standard
deviation based on the individual age determinations; for those
clusters with clearly discrepant values, we also give these numbers
excluding these discrepant determinations. In addition, we have also
included those cluster for which we determined ages of log( Age/yr )
$\le 6.6$ but for which no strong H$\alpha$ EW measurements were
provided.

\begin{table*}
\caption[ ]{\label{ha-ages.tab}Age comparison for the individual
clusters in our Antennae cluster sample with EW$_{{\rm H}\alpha} >
1.0$\AA.\\
Values in italics are clearly discrepant values.}
{\scriptsize
\begin{center}
\begin{tabular}{cccccccccrl}
\hline
\hline
\multicolumn{1}{c}{ID} & \multicolumn{1}{c}{EW$_{{\rm H}\alpha}$} &
\multicolumn{1}{c}{\it Q-Q} & \multicolumn{1}{c}{BB+H$\alpha$} &
\multicolumn{1}{c}{AnalySED; {\it Z} free} &
\multicolumn{1}{c}{AnalySED; $Z_\odot$} & \multicolumn{1}{c}{3DEF} &
\multicolumn{1}{c}{O/IR} && \multicolumn{2}{c}{log( Age/yr )}\\
\cline{3-8}\cline{10-11}
& \multicolumn{1}{c}{(\AA)} & \multicolumn{6}{c}{log( Age/yr )} &&
\multicolumn{1}{c}{Mean} & \multicolumn{1}{c}{$\sigma$} \\
\hline
G2-04 & 2.9   &  6.5  & 6.7    &  7.1 &  6.9 &  6.8   &  7.2 && 6.87 & 0.24 \\
G2-05 & 2.5   &{\it 7.5}& 6.8  &  7.1 &  6.6 &  6.8   &  {\it 7.6} && 6.98 & 0.29 \\
      &       &       &        &      &      &        &      &&(6.83 & 0.18)$^b$ \\
G2-07 & 1.3   &  6.5  & 6.5    &  7.1 &  6.6 &  6.8   &  7.1 && 6.77 & 0.26 \\
G2-08 & 2.9   &  6.5  & 6.8    &  7.2 &  6.6 &  6.8   &  7.0 && 6.82 & 0.23 \\
G2-10 & 1.4   &  6.6  & 6.8    &  6.9 &  6.9 &  6.8   &  7.4 && 6.90 & 0.24 \\
G2-12 & 2.5   &  6.6  & 6.8    &  7.3 &  6.6 &  6.8   &  6.8 && 6.82 & 0.23 \\
G2-14 & 3.7   &  6.5  & 6.1    &  6.9 &  6.6 &{\it 9.0}&{\it 9.3} && 7.40 & 1.26 \\
      &       &       &        &      &      &        &      &&(6.53 & 0.29)$^b$ \\
G2-15 & 3.5   &  6.5  & 6.5    &  7.1 &  6.6 &  6.6   &  6.8 && 6.68 & 0.21 \\
\\
\multicolumn{2}{l}{Others with}              \\
\multicolumn{2}{l}{log( Age/yr ) $\le 6.6$,} & G2-17 & G2-19 & G2-17$^a$ & G2-09, & -- & G2-17$^a$ \\
\multicolumn{2}{l}{but no H$\alpha$}         &       &       &           & G2-19 \\
\hline
\end{tabular}
\end{center}
{\sc Notes:} $^a$ Here, we used a limit of log( Age/yr ) $\le 6.9$
because of the offset introduced by the age-metallicity degeneracy;
$^b$ After exclusion of the values in italic font.\hfill}
\end{table*}

One can see immediately that the correspondence between the age
estimates for a given cluster among the different analysis approaches
is close. We note that the age estimates obtained using the sequential
O/IR and AnalySED approaches, with metallicity as a free parameter are
offset with respect to most of the other methods. This simply reflects
the age-metallicity degeneracy for these young ages. Nevertheless, the
fact that we obtain very similar age estimates using a variety of
independent approaches that may or may not include H$\alpha$
luminosities as an additional contraint is reassuring. In principle,
this validates the different approaches used here: nearly all clusters
that should be young based on the presence of strong H$\alpha$
emission are indeed young (with between 0 and 2 outliers for all
methods), and the ones predicted to be young based on these methods
all have H$\alpha$ (except for, again, between 0 and 2 outliers). The
outliers are characterised by the largest error bars, which is an
additional argument in support of the robustness of the variety of
models employed here. Unfortunately, there are no young extragalactic
star clusters available in the current literature for which
independent age estimates have been obtained via either spectroscopy
or detailed analysis of their resolved colour-magnitude diagrams; the
use of H$\alpha$ EWs, as done here, is therefore the closest we can
get to an independent validation of our approach.

As an example of the uncertainties inherent to the use of broad-band
(and H$\alpha$) fluxes to obtain the individual cluster ages, we
direct the reader's attention to cluster G2-14, for which we found the
most discrepant age estimates among the variety of approaches
used. This cluster shows strong H$\alpha$ emission, and must therefore
be young. Nevertheless, two of the approaches employed assigned this
cluster an age of greater than 1 Gyr. Upon close inspection, those
discrepant estimates originated from the two approaches that
essentially leave the metallicity as a free parameter. This hints at
an origin related to the age-metallicity degeneracy. However, we also
point out that the error bars assigned by the AnalySED approach are
among the largest in our cluster sample. Statistically, the AnalySED
age estimate (with metallicity as a free parameter) is therefore also
consistent with a young age.

On the other hand, {\it all} of the different approaches employed
here, for instance, estimate the age of cluster G2-17 to be in the
range $6.6 \lesssim \log( {\rm Age/yr} ) \lesssim 6.9$ (see the
Appendix for details), despite not having strong H$\alpha$ emission.
In spite of this, the uniformity of this result among the different
methods is encouraging; it implies a high degree of reproducibility.

\section{Filter system conversions}
\label{filters.sec}

Largely owing to the unavailability of SSP models computed for the
{\sl HST}-flight system magnitudes, many of the early studies based on
{\sl HST} imaging observations of extragalactic star cluster systems
used the equations given by Holtzman et al. (1995) to convert {\sl
HST} {\sc stmag} magnitudes to the ``standard'' Johnson-Cousins
system. Despite recent updates of many of the leading SSP models,
which now include theoretical magnitudes in the {\sc stmag} system,
many workers in the field, including ourselves in this paper (see our
photometry of the Antennae clusters), continue to use the Holtzman et
al. (1995) conversions.

It is clear, however, that any conversion based on generic spectral
properties of a given stellar population will introduce biases and
additional uncertainties that could, in principle, be avoided by
retaining one's photometry in the filter system used for the
observations. In this section, we explore the extent of these
additional uncertainties by comparing our model fitting results for
the full NGC 3310 cluster sample, presented in de Grijs et
al. (2003c), based on both the original {\sl HST}/WFPC2 photometry and
on the transformed magnitudes in the Landolt (KPNO) $UBVRI$ system
used by Holtzman et al. (1995).

In order to do so, we folded the {\sc galev} SSP models through these
filter transmission curves, which were kindly made available by Jon
Holtzman (priv. comm.). Despite being a standard system, the
appropriate filter transmission curves have not been published in
their entirety. We transformed our F336W, F439W, F606W and F814W
magnitudes using Eqs. (6)--(9) below (for the WF3 chip of the WFPC2
camera), using an iterative approach; following Holtzman et al.'s
(1995) recommendations, we did not transform the F300W magnitudes, but
used the {\sl HST} flight system magnitude for this part of the
broad-band cluster SEDs. The conversions from the F336W, F439W, and
F814W filters to their $UBI$ counterparts are based on the
observational transformations from the WFPC2 flight system (Holtzman
et al.'s Table 7), while the transformation of the F606W magnitudes to
the $V$ filter relies on the conversion of the {\it synthetic} WFPC2
system to the standard $V$ band (their Table 10).

\begin{eqnarray}
U =\hspace{-0.6cm}& &-2.5 \times \log(\mbox{counts s}^{-1}) - 0.240 \times (U-V) + \hfill \nonumber \\
   & &0.048 \times (U-V)^2 + 18.764 + 2.5 \times \log( 2.003 ) \\
B =\hspace{-0.6cm}& &-2.5 \times \log(\mbox{counts s}^{-1}) + 0.003 \times (B-V) + \hfill \nonumber \\
   & &-0.088 \times (B-V)^2 + 20.070 + 2.5 \times \log( 2.003 ) \\
V =\hspace{-0.6cm}& &-2.5 \times \log(\mbox{counts s}^{-1}) + 0.254 \times (V-I) + \hfill \nonumber \\
   & &+0.012 \times (V-I)^2 + 22.093 + 2.5 \times \log( 2.003 ) \\
& &(\mbox{for }(V-I) \le 2.0) \nonumber \\
I =\hspace{-0.6cm}& &-2.5 \times \log(\mbox{counts s}^{-1}) - 0.062 \times (V-I) + \hfill \nonumber \\
   & &0.025 \times (V-I)^2 + 20.839 + 2.5 \times \log( 2.003 )
\end{eqnarray}

As before, we will first discuss the characteristics (i.e.,
predominantly the mean and spread) of the age and mass distributions
of the entire cluster population, and conclude with a one-to-one
comparison of the results obtained for the individual clusters. For
the analysis presented in this section, we have adopted the AnalySED
multi-dimensional modelling approach, using the {\sc galev} SSP
models. We attempted to obtain the cluster ages and masses under three
sets of assumptions: {\it (i)} unrestricted fits, i.e., we left all of
the cluster ages, masses, metallicities and extinction values as free
parameters; {\it (ii)} as for {\it (i)}, but assuming solar
metallicity for all clusters; {\it (iii)} as for {\it (ii)}, but now
also assuming a generic (arbitrarily low) extinction value for each
cluster of E$(B-V) = 0.1$ mag.

\begin{figure}
\psfig{figure=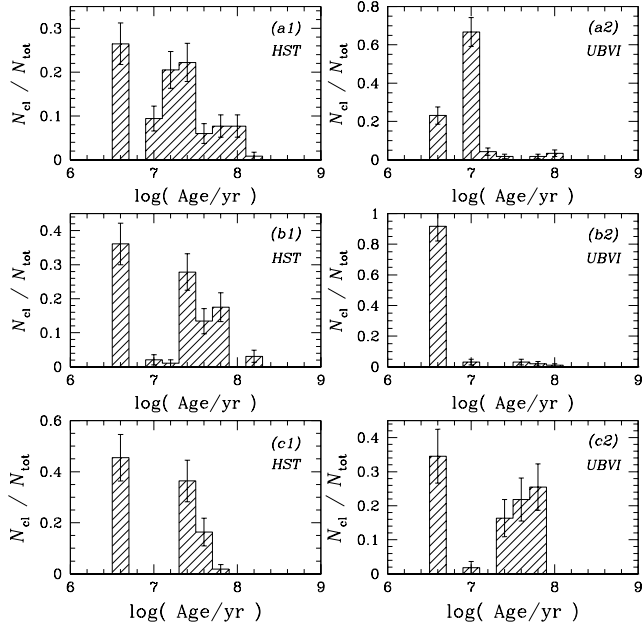,width=9cm}
\caption[]{Comparison of the relative age distribution of the NGC 3310
clusters, based on the {\sl HST} flight system photometry (left-hand
column, panels a1, b1, and c1) and on the converted $UBVI$ magnitudes
(right-hand column, panels a2, b2, and c2). Panels a1 and a2 are based
on model fits in which all of the cluster ages, masses, metallicities
and extinction values were left as free parameters; in panels b1 and
b2 we restricted the fits to solar metallicity, and in panels c1 and
c2 we also adopted a generic extinction of E$(B-V) = 0.1$ mag.}
\label{convages.fig}
\end{figure}

In Fig. \ref{convages.fig} we present the results for the relative age
distributions, in the left-hand column using the {\sl HST}-flight
system magnitudes as our basis, and in the right-hand column using the
transformed F300W-{\it UBVI} photometry. From top to bottom, the fits
become more and more restricted, following the assumptions laid out
above. Despite this being the same sample as analysed in de Grijs et
al. (2003a,c), the resulting age distribution in panel (a) is
different from that published previously. The main reason for this
difference is that the fits discussed in this section do not include
the NIR passbands, so that we are less sensitive to metallicity
variations (Anders et al. 2004b). In other words, the strong peak seen
at log(Age/yr) $\sim 6.6$ is caused by the age-metallicity effect, and
also by the fact that the youngest age in our models corresponds to
log(Age/yr) = 6.6 (so that younger clusters will automatically be
assigned the minimum age in the models). This applies to the strong
peaks seen at this age in all of the panels of
Fig. \ref{convages.fig}. We note that the overall metallicity of NGC
3310 is known to be significantly subsolar (see Section
\ref{n3310cov.sec}), but we will use the assumption of solar
metallicity in this section to emphasize a number of technical
concerns relevant for similar studies in this field.

Apart from this obvious signature of the age-metallicity degeneracy,
the overall relative age distribution, i.e., the mean value and its
spread, based on the {\sl HST} flight system photometry is retrieved
relatively robustly under the various assumptions employed (see Fig.
\ref{convages.fig}, left-hand panels); the range of ages found for the
NGC 3310 clusters is consistent with independently determined age
estimates for the star-forming regions in this galaxy (see de Grijs et
al. 2003a,c for a detailed discussion). However, if we now examine the
age distributions based on the transformed F300W-{\it UBVI}
magnitudes, we see that {\it (i)} our results are severely affected by
the age-metallicity degeneracy, and {\it (ii)} that the ``transformed
ages'' do not correspond to even remotely similar ages as obtained
from the {\sl HST}-flight magnitudes. Only by severely restricting our
model fits (Fig. \ref{convages.fig}c2) are we able to retrieve a
similar age distribution as we obtained from the direct match of our
SSP model grid to the {\sl HST}-flight system magnitudes. This
provides, therefore, a very strong argument against using photometry
based on even robust filter conversions; the effect of such
transformations is the unavoidable introduction of biases and
additional uncertainties, which thus makes the various fitting
routines less reliable and robust.

\begin{figure}
\psfig{figure=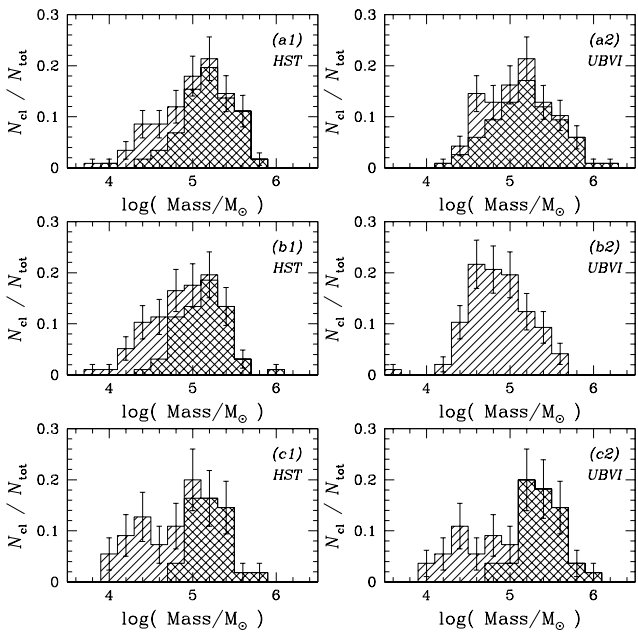,width=9cm}
\caption[]{Comparison of the relative mass distribution of the NGC
3310 clusters, based on the {\sl HST} flight system photometry
(left-hand column, panels a1, b1, and c1) and on the converted $UBVI$
magnitudes (right-hand column, panels a2, b2, and c2). Panels a1 and
a2 are based on model fits in which all of the cluster ages, masses,
metallicities and extinction values were left as free parameters; in
panels b1 and b2 we restricted the fits to solar metallicity, and in
panels c1 and c2 we also adopted a generic extinction of E$(B-V) =
0.1$ mag. The cross-hatched distributions contain clusters
characterised by log( Age/yr ) $> 6.8$, to avoid the effects of the
age-metallicity degeneracy (see text).}
\label{convmasses.fig}
\end{figure}

If we now consider the resulting mass distributions, shown in
Fig. \ref{convmasses.fig}, we see -- somewhat to our surprise -- that
to first order these can be reproduced relatively robustly, based on
either set of filter transmission curves. This is owing to the fact
that our mass estimates are predominantly determined by a global
scaling between the entire observed SED and the most appropriate model
SED, rather than on the exact shape of the SED.

To highlight the robustness with which we can retrieve the global
characteristics of the mass distribution, and also to address the
effects caused by the age-metallicity degeneracy, we show the relative
mass distributions for only the clusters with ages, log(Age/yr) $>
6.8$ as the cross-hatched histograms, where relevant. The global
characteristics, including the mean and spread ($\sigma$), of all mass
distributions shown in Fig. \ref{convmasses.fig} are listed in Table
\ref{transformed.tab}. We note that while the global characteristics
of the mass distributions are reproduced robustly, the details of the
distributions differ among the panels of Fig. \ref{convmasses.fig}. To
illustrate this, we applied KS tests to the relevant data sets, of
which the results are summarised in Table \ref{kstests.tab}.

\begin{table*}
\caption[ ]{\label{transformed.tab}Characteristics of the overall
relative mass distributions of the NGC 3310 cluster sample, as
discussed w.r.t. Fig. \ref{convmasses.fig},
respectively.}
{\scriptsize
\begin{center}
\begin{tabular}{cccccccc}
\hline
\hline
\multicolumn{1}{c}{System} & \multicolumn{1}{c}{Selection} &
\multicolumn{1}{c}{Restrictions} &
\multicolumn{2}{c}{log(Mass/M$_\odot$)} \\
\cline{4-5}
& & & \multicolumn{1}{c}{Mean} & \multicolumn{1}{c}{$\sigma$} \\
\hline
{\sl HST}  & all                 & unrestricted & 5.03 & 0.41 \\
{\sl HST}  & log(Age/yr) $> 6.8$ & unrestricted & 5.17 & 0.33 \\
{\sl HST}  & all                 & $Z_\odot$    & 4.92 & 0.44 \\
{\sl HST}  & log(Age/yr) $> 6.8$ & $Z_\odot$    & 5.14 & 0.29 \\
{\sl HST}  & all                 & $Z_\odot$, E$(B-V) = 0.1$ & 4.87 & 0.46 \\
{\sl HST}  & log(Age/yr) $> 6.8$ & $Z_\odot$, E$(B-V) = 0.1$ & 5.20 & 0.24 \\
\\
{\it UBVI} & all                 & unrestricted & 5.11 & 0.40 \\
{\it UBVI} & log(Age/yr) $> 6.8$ & unrestricted & 5.18 & 0.39 \\
{\it UBVI} & all                 & $Z_\odot$    & 4.87 & 0.36 \\
{\it UBVI} & all                 & $Z_\odot$, E$(B-V) = 0.1$ & 5.08 & 0.50 \\
{\it UBVI} & log(Age/yr) $> 6.8$ & $Z_\odot$, E$(B-V) = 0.1$ & 5.38 & 0.26 \\
\hline
\end{tabular}
\end{center}
}
\end{table*}

\begin{table}
\caption[ ]{\label{kstests.tab}Detailed comparison of the mass
distributions in Fig. \ref{convmasses.fig}. Probabilities refer to the
chances that both the ``{\sl HST}'' and the ``{\sl UBVI}'' samples were
drawn from the same distribution, based on KS statistics.}
{\scriptsize
\begin{center}
\begin{tabular}{ccc}
\hline
\hline
\multicolumn{1}{c}{Selection} &
\multicolumn{1}{c}{Restrictions} &
\multicolumn{1}{c}{Probability} \\
\hline
all                 & unrestricted & 0.663 \\
log(Age/yr) $> 6.8$ & unrestricted & 0.379 \\
all                 & $Z_\odot$    & 0.250 \\
all                 & $Z_\odot$, E$(B-V) = 0.1$ & 0.047 \\
log(Age/yr) $> 6.8$ & $Z_\odot$, E$(B-V) = 0.1$ & 0.015 \\
\hline
\end{tabular}
\end{center}
}
\end{table}

\begin{figure}
\psfig{figure=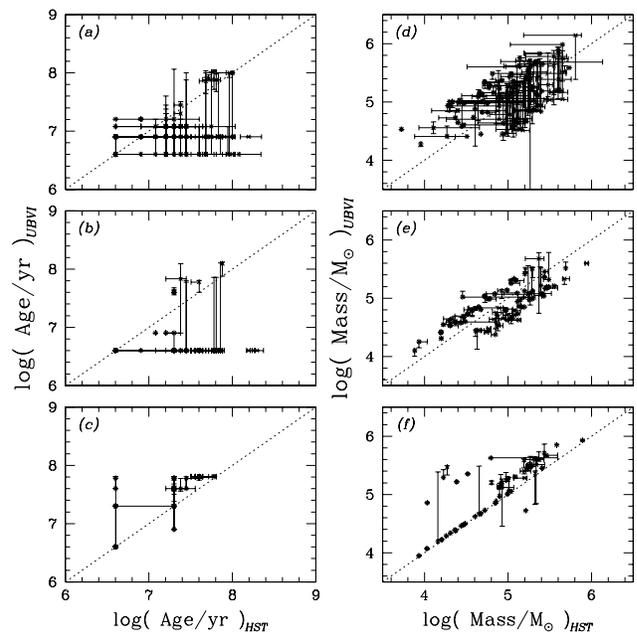,width=9cm}
\caption[]{One-to-one comparisons between the age (left) and mass
(right) estimates of the NGC 3310 clusters, based on the {\sl HST}
flight system photometry vs. the converted $UBVI$ magnitudes. The
dotted lines are the loci of equality.}
\label{convall.fig}
\end{figure}

Finally, we compare the individual cluster age and mass estimates
obtained from both sets of filter transmission curves, and all three
sets of assumptions in Fig. \ref{convall.fig}. The individual panels
in this figure reflect the discussion above: the individual age
estimates are severely discrepant, while the mass estimates are
relatively robust from one cluster to another. We note the existence
of two ``sequences'' in Figs. \ref{convall.fig}d and e. These are
indeed caused by the age-metallicity degeneracy discussed above; the
upper (``left-hand'') sequence consists predominantly of clusters that
are found in the strong peak at our minimum age limit.

Thus, we conclude from this analysis that the overall characteristics
of a cluster system's relative mass distribution, and to some extent
the individual cluster mass estimates as well, can be reproduced
fairly robustly under a variety of relevant fitting assumptions and
conversions between filter systems. However, in order to derive more
robust age estimates, one should ideally retain one's photometry in
the filter system used for the observations. This is in order to
introduce as few biases and additional uncertainties as possible. We
emphasize that we have followed this route in previous publications in
which we applied these results.

\section{Summary and Conclusions}
\label{final.sec}

The increasing availability of high-resolution {\sl HST} imaging
observations across a wide wavelength range (from the mid-UV to the
NIR) has revolutionised studies of extragalactic star cluster
populations.

The age, mass, metallicity and extinction of unresolved extragalactic
star clusters can be derived from the spectral energy distributions
measured in broad-band photometric systems, by comparing them with
cluster evolution models. This method is applied in the literature
based on observations with different sets of photometric filters and
different sets of cluster evolution models.

In this paper, we investigated the accuracy of these determinations by
comparing the parameters derived from different filter sets and
different sets of cluster models. To this purpose, we used the $UBVI$
and {\sl HST} photometry of 20 clusters in the Antennae galaxies and
17 clusters in NGC 3310, analysed in the different ways that are found
in the literature. We assess the systematic uncertainties in age and
mass determinations, and to a lesser extent also in extinction and
metallicity determinations. We compare the results with those of the
extensively tested and well-validated AnalySED approach (see Sections
\ref{method5.sec} and \ref{artdata.sec}). The results, which are
summarised below, give us a handle on the systematic uncertainties one
needs to contend with when comparing results obtained by different
groups, each using a different modelling technique.

We first examined the model parameters used by the various groups that
contributed to the final results. We concluded that the differences
among the various SSP models, specifically among the various Bruzual
\& Charlot model incarnations, the Starburst99 and the {\sc galev} SSP
models, are random and do not bias the results in a systematic
fashion. Similarly, the variety of (foreground) extinction curves used
by the various groups show minimal differences over the wavelength
range considered in this paper (from the mid-UV to the NIR), and any
differences are deemed unimportant in view of the photometric
uncertainties of similar magnitude. Thus, we conclude that any
significant differences among the resulting parameters are due to the
details of the various methods used, rather than to the models
themselves (see also Yi 2003).

The methods used in this study are: 
\begin{enumerate} 
\item the multi-dimensional SED analysis (Anders et al. 2004b) and
(Bik et al. 2003); 
\item the broad-band fluxes with H$\alpha$ method (Whitmore \& Zhang
2002); 
\item the reddening-free parameter method ($Q-Q$) (Whitmore et al. 
1999), and
\item the optical/NIR sequential analysis method (Parmentier et
al. 2003).
\end{enumerate}

The free parameters in the matching of cluster models to an observed
photometric cluster SED are, mass, age, metallicity and extinction.

The effects of (foreground) extinction and metallicity variations are
small, within a given cluster sample. However, on an individual
cluster basis, it is very difficult to estimate a best-fitting
metallicity, despite major efforts and recent improvements made in the
modelling techniques (e.g., de Grijs et al. 2003a,b, Anders et
al. 2004b). This agrees with the conclusion by Bastian et al. (2004)
that they could not determine the metallicity of individual clusters
in the interacting spiral galaxy M51, but that the derived age and
mass distributions are only slightly affected compared to the
solar-metallicity case. Thus, the effect of metallicity variations is
minimal in view of the uncertainties, and conclusions regarding
cluster age and mass distributions appear robust.

For this reason, and also in order to compare our results to
previously published results, we adopted solar metallicities for most
of the modelling done in this paper. We find that the ages of a
cluster, measured using different methods, can differ drastically, but
that the ``age distribution'' of a given young ($\lesssim 10^9$ yr)
cluster system derived using the different methods for a fixed
metallicity shows the same main features. We also find that the
differences in the mass distributions, derived using different
approaches, are much smaller than those in the age
distributions. Thus, the mass distributions can be obtained with a
higher degree of confidence than the age distributions. We determine
accuracies for our age and mass distributions, based on the
simplifying assumption that these properties are roughly following a
Gaussian distribution, of $\sigma_t \equiv \Delta \langle \log( {\rm
Age / yr} ) \rangle \le 0.35$ and $\sigma_M \equiv \Delta \langle
\log( M_{\rm cl} / {\rm M}_\odot ) \rangle \le 0.14$,
respectively. While this assumption may not hold in general, the
differences resulting from its blanket application to all results
presented in this paper are indicative of the degree of confidence we
can attribute to our fits. The very small spread in the mean mass
shows that, if there is a peak in the mass distribution, the retrieval
of its absolute value is relatively insensitive to the approach taken.

Since the actual age distributions of our cluster samples are
distinctly not unimodal, the use of a ``mean'' value and ``width'' for
the comparison of methods provides a straightforward, yet not a
physically interesting criterion. The absolute age distribution is
rather sensitive to the adopted method, and one should therefore be
very cautious when comparing such values among different studies. As
we showed previously (e.g., de Grijs et al. 2001, 2003a,b,c, Anders et
al. 2004a,b), the {\it relative} age {\it distributions} (i.e., the
presence or absence of peaks) can be retrieved with a very high level
of confidence, however.

It is important to keep in mind that all of the techniques employed in
this paper are affected by various selection effects and artifacts
that require careful consideration when comparing results. For
instance, the peak at young ages in the age distribution of
Fig. \ref{convmasses.fig} is caused by the use of a model of this age
as the youngest time-step, while the bimodality in a number of the
resulting age distributions is due to differences in the treatment of
RSG stars (see also Whitmore \& Zhang 2002). Nevertheless, the close
correspondence among the resulting cluster parameters based on the
variety of methods used is reassuring in the context of the robustness
of our parameter determination techniques.

We note that as extensive a wavelength coverage as possible is
required to obtain robust age and mass estimates for the individual
objects, with reasonable uncertainties. We also show that the
conversion of the {\sl HST} photometry to the ``standard'' Johnson and
Cousins $UBVI$ photometry introduces an extra uncertainty in the
determination of the ages and masses of the clusters, because the
calibration of this conversion is based on stars rather than clusters.

\section*{Acknowledgments} The idea for the comparison among methods and
models carried out in this paper originated in the stimulating
scientific atmosphere of the August 2002 ESO workshop on
``Extragalactic Globular Cluster Systems'', organised by Markus
Kissler-Patig and colleagues. We are therefore indebted to the LOC and
SOC of this meeting. MS thanks the workshop organisation for the
financial assistance that enabled her to attend, and the Russian
Foundation for Basic Research for travel grant 02-02-27032. We also
acknowledge useful discussions with Brad Whitmore (and for providing
us with the Antennae data), Rupali Chandar (who also assisted us
significantly with the BB+H$\alpha$ analysis), Nicolai Bissantz, and
Jon Holtzman, to whom we are also grateful for providing us with
unpublished filter transmission curves. RdeG acknowledges an equipment
grant from the Royal Society. RdeG, PA and HL acknowledge hospitality
of the International Space Science Institute in Bern, Switzerland,
during the final stage of this project. This paper is based on
archival observations with the NASA/ESA {\sl Hubble Space Telescope},
obtained at the Space Telescope Science Institute, which is operated
by the (US) Association of Universities for Research in Astronomy
(AURA), Inc., under NASA contract NAS 5-26555.

%\vfill\eject
\appendix

\section{Results from the individual approaches}

In this Appendix, we present the best-fitting results for the
individual clusters as obtained by applying the different methods
discussed in this paper to the cluster photometry presented in Section
\ref{data.sec}. The following tables are arranged following the order
of the age and mass histograms in
Figs. \ref{n3310ages.fig}--\ref{n4038masses.fig}.

\begin{table*}
\caption[ ]{Results for the NGC 3310 clusters from the
multi-dimensional SED fits, using the AnalySED approach with cluster
ages, masses, metallicities and extinction values as free parameters
(cf. Figs. \ref{n3310ages.fig}a and \ref{n3310masses.fig}a).}
{\scriptsize
\begin{center}
\begin{tabular}{ccccccccccrrcccc}
\hline
\hline
\multicolumn{1}{c}{ID} & \multicolumn{3}{c}{$Z$} &&
\multicolumn{3}{c}{E$(B-V)$ (mag)} && \multicolumn{3}{c}{Age ($\times
10^7$ yr)} && \multicolumn{3}{c}{Mass ($\times 10^5$ M$_\odot$)}\\
\cline{2-4}\cline{6-8}\cline{10-12}\cline{14-16}
& \multicolumn{1}{c}{Min.} & \multicolumn{1}{c}{Best} &
\multicolumn{1}{c}{Max.} && \multicolumn{1}{c}{Min.} &
\multicolumn{1}{c}{Best} & \multicolumn{1}{c}{Max.} &&
\multicolumn{1}{c}{Min.} & \multicolumn{1}{c}{Best} &
\multicolumn{1}{c}{Max.} && \multicolumn{1}{c}{Min.} &
\multicolumn{1}{c}{Best} & \multicolumn{1}{c}{Max.} \\
\hline
G1-01 & 0.0003 & 0.0004 & 0.0005 && 0.17 & 0.25 & 0.32 && 0.84 & 1.20 & 1.56 && 1.88 & 2.69 & 3.50 \\
G1-02 & 0.0004 & 0.0004 & 0.0004 && 0.00 & 0.00 & 0.10 && 1.20 & 1.60 & 1.60 && 1.95 & 1.95 & 2.23 \\
G1-03 & 0.0040 & 0.0080 & 0.0080 && 0.00 & 0.00 & 0.15 && 1.60 & 3.60 & 4.80 && 0.55 & 0.73 & 0.94 \\
G1-04 & 0.0200 & 0.0200 & 0.0200 && 0.00 & 0.05 & 0.05 && 1.20 & 2.40 & 2.80 && 0.27 & 0.66 & 0.77 \\
G1-05 & 0.0004 & 0.0040 & 0.0080 && 0.00 & 0.10 & 0.30 && 2.80 & 6.00 &12.40 && 0.78 & 1.07 & 1.83 \\
G1-06 & 0.0003 & 0.0004 & 0.0005 && 0.07 & 0.10 & 0.13 && 0.84 & 1.20 & 1.56 && 1.40 & 2.00 & 2.60 \\
G1-07 & 0.0004 & 0.0004 & 0.0200 && 0.20 & 0.20 & 0.25 && 0.40 & 0.80 & 0.80 && 2.67 & 5.62 & 5.62 \\
G1-08 & 0.0004 & 0.0004 & 0.0004 && 0.10 & 0.10 & 0.15 && 6.80 &10.00 &10.00 && 0.69 & 0.80 & 0.90 \\
G1-09 & 0.0200 & 0.0500 & 0.0500 && 0.00 & 0.00 & 0.00 && 2.80 & 4.80 & 6.80 && 0.40 & 0.58 & 0.70 \\
G1-10 & 0.0004 & 0.0004 & 0.0040 && 0.10 & 0.25 & 0.25 && 0.80 & 1.20 & 1.20 && 0.61 & 2.80 & 2.80 \\
G1-11 & 0.0040 & 0.0040 & 0.0200 && 0.00 & 0.15 & 0.25 && 1.20 & 2.00 & 4.80 && 0.61 & 0.99 & 1.45 \\
G1-12 & 0.0040 & 0.0040 & 0.0080 && 0.00 & 0.05 & 0.05 && 1.60 & 1.60 & 1.60 && 1.77 & 2.49 & 2.49 \\
G1-13 & 0.0080 & 0.0500 & 0.0500 && 0.00 & 0.00 & 0.05 && 4.40 & 5.20 &12.00 && 0.74 & 0.76 & 1.10 \\
G1-14 & 0.0040 & 0.0200 & 0.0200 && 0.00 & 0.15 & 0.35 && 1.20 & 4.80 &23.20 && 0.49 & 0.89 & 1.31 \\
G1-15 & 0.0004 & 0.0080 & 0.0200 && 0.00 & 0.10 & 0.30 && 0.80 & 5.60 &10.80 && 0.24 & 1.20 & 1.87 \\
G1-16 & 0.0004 & 0.0004 & 0.0200 && 0.00 & 0.55 & 0.65 && 0.80 & 1.60 &60.40 && 0.17 & 2.11 & 3.10 \\
G1-17 & 0.0200 & 0.0500 & 0.0500 && 0.00 & 0.00 & 0.00 && 5.60 & 5.60 &13.20 && 1.00 & 1.01 & 1.60 \\
\hline
\end{tabular}
\end{center}
}
\end{table*}

\begin{table*}
\caption[ ]{Results for the NGC 3310 clusters from the
multi-dimensional SED fits, using the AnalySED approach with cluster
ages, masses, and extinction values as free parameters, and adopting
solar metallicity for all clusters (cf. Figs. \ref{n3310ages.fig}b and
\ref{n3310masses.fig}b).}  {\scriptsize
\begin{center}
\begin{tabular}{ccccccrrcccc}
\hline
\hline
\multicolumn{1}{c}{ID} & \multicolumn{3}{c}{E$(B-V)$ (mag)} &&
\multicolumn{3}{c}{Age ($\times 10^7$ yr)} && \multicolumn{3}{c}{Mass
($\times 10^5$ M$_\odot$)}\\ 
\cline{2-4}\cline{6-8}\cline{10-12}
& \multicolumn{1}{c}{Min.} & \multicolumn{1}{c}{Best} &
\multicolumn{1}{c}{Max.} && \multicolumn{1}{c}{Min.} &
\multicolumn{1}{c}{Best} & \multicolumn{1}{c}{Max.} &&
\multicolumn{1}{c}{Min.} & \multicolumn{1}{c}{Best} &
\multicolumn{1}{c}{Max.} \\
\hline
G1-01 & 0.17 & 0.25 & 0.32 && 0.28 & 0.40 & 0.52 && 0.47 & 0.67 & 0.87 \\
G1-02 & 0.07 & 0.10 & 0.13 && 0.28 & 0.40 & 0.52 && 0.39 & 0.55 & 0.72 \\
G1-03 & 0.00 & 0.00 & 0.00 && 5.20 & 5.60 & 5.60 && 1.09 & 1.17 & 1.17 \\
G1-04 & 0.00 & 0.00 & 0.05 && 0.80 & 0.80 & 2.00 && 0.27 & 0.27 & 0.67 \\
G1-05 & 0.05 & 0.05 & 0.10 && 4.00 & 6.00 & 6.00 && 0.75 & 1.03 & 1.17 \\
G1-06 & 0.10 & 0.15 & 0.19 && 0.28 & 0.40 & 0.52 && 0.41 & 0.58 & 0.76 \\
G1-07 & 0.17 & 0.25 & 0.32 && 0.28 & 0.40 & 0.52 && 1.87 & 2.67 & 3.47 \\
G1-08 & 0.35 & 0.40 & 0.40 && 0.40 & 0.40 & 0.40 && 0.13 & 0.16 & 0.16 \\
G1-09 & 0.00 & 0.00 & 0.00 && 6.40 & 6.40 & 6.80 && 0.66 & 0.68 & 0.68 \\
G1-10 & 0.21 & 0.30 & 0.39 && 0.28 & 0.40 & 0.52 && 0.57 & 0.82 & 1.06 \\
G1-11 & 0.00 & 0.00 & 0.00 && 2.80 & 3.20 & 4.40 && 0.81 & 0.88 & 1.09 \\
G1-12 & 0.17 & 0.25 & 0.32 && 0.28 & 0.40 & 0.52 && 0.92 & 1.31 & 1.70 \\
G1-13 & 0.00 & 0.00 & 0.00 && 7.20 & 7.60 & 7.60 && 0.88 & 0.88 & 0.88 \\
G1-14 & 0.10 & 0.15 & 0.15 && 3.60 & 5.60 & 6.40 && 0.79 & 1.14 & 1.20 \\
G1-15 & 0.00 & 0.05 & 0.05 && 4.40 & 5.60 & 6.00 && 0.82 & 1.09 & 1.14 \\
G1-16 & 0.00 & 0.00 & 0.25 && 5.20 &30.00 &48.80 && 1.08 & 1.56 & 2.44 \\
G1-17 & 0.00 & 0.00 & 0.00 && 7.20 & 7.60 & 8.40 && 1.27 & 1.28 & 1.32 \\
\hline
\end{tabular}
\end{center}
}
\end{table*}

\begin{table*}
\caption[ ]{Results for the NGC 3310 clusters from the 3DEF approach,
using the {\sc galev} SSP models and adopting solar metallicity for
all clusters (cf. Figs. \ref{n3310ages.fig}c and
\ref{n3310masses.fig}c).}
{\scriptsize
\begin{center}
\begin{tabular}{cccccccccccc}
\hline
\hline
\multicolumn{1}{c}{ID} & \multicolumn{3}{c}{E$(B-V)$ (mag)} &&
\multicolumn{3}{c}{log( Age/yr )} && \multicolumn{3}{c}{log(
Mass/M$_\odot$ )}\\
\cline{2-4}\cline{6-8}\cline{10-12}
& \multicolumn{1}{c}{Min.} & \multicolumn{1}{c}{Best} &
\multicolumn{1}{c}{Max.} && \multicolumn{1}{c}{Min.} &
\multicolumn{1}{c}{Best} & \multicolumn{1}{c}{Max.} &&
\multicolumn{1}{c}{Min.} & \multicolumn{1}{c}{Best} &
\multicolumn{1}{c}{Max.} \\
\hline
G1-01 & 0.22 & 0.26 & 0.32 && 6.60 & 6.60 & 6.60 && 4.65 & 4.70 & 4.78 \\
G1-02 & 0.06 & 0.12 & 0.16 && 6.60 & 6.60 & 6.60 && 4.62 & 4.69 & 4.74 \\
G1-03 & 0.00 & 0.00 & 0.04 && 7.51 & 7.78 & 7.81 && 4.81 & 5.02 & 5.07 \\
G1-04 & 0.00 & 0.04 & 0.16 && 7.08 & 7.08 & 7.56 && 4.41 & 4.45 & 5.02 \\
G1-05 & 0.00 & 0.10 & 0.18 && 6.90 & 7.78 & 7.86 && 4.06 & 4.98 & 5.05 \\
G1-06 & 0.06 & 0.12 & 0.18 && 6.60 & 6.60 & 6.60 && 4.54 & 4.63 & 4.71 \\
G1-07 & 0.24 & 0.28 & 0.32 && 6.60 & 6.60 & 6.60 && 5.29 & 5.34 & 5.38 \\
G1-08 & 0.00 & 0.00 & 0.10 && 6.90 & 7.81 & 7.86 && 3.70 & 4.60 & 4.72 \\
G1-09 & 0.00 & 0.00 & 0.10 && 7.38 & 7.83 & 7.88 && 4.50 & 4.82 & 4.92 \\
G1-10 & 0.26 & 0.32 & 0.36 && 6.60 & 6.60 & 6.60 && 4.74 & 4.81 & 4.86 \\
G1-11 & 0.00 & 0.06 & 0.12 && 6.90 & 7.60 & 7.86 && 4.24 & 5.03 & 5.25 \\
G1-12 & 0.22 & 0.28 & 0.34 && 6.60 & 6.60 & 6.60 && 4.97 & 5.04 & 5.10 \\
G1-13 & 0.00 & 0.00 & 0.14 && 7.45 & 7.88 & 8.13 && 4.71 & 4.95 & 5.08 \\
G1-14 & 0.00 & 0.18 & 0.30 && 6.90 & 7.56 & 8.43 && 4.11 & 4.84 & 5.22 \\
G1-15 & 0.00 & 0.08 & 0.18 && 6.90 & 7.81 & 7.88 && 4.06 & 5.05 & 5.16 \\
G1-16 & 0.00 & 0.30 & 0.48 && 6.90 & 7.81 & 8.95 && 4.13 & 5.09 & 5.51 \\
G1-17 & 0.00 & 0.00 & 0.12 && 7.86 & 8.09 & 8.25 && 5.07 & 5.17 & 5.28 \\
\hline
\end{tabular}
\end{center}
}
\end{table*}

\begin{table*}
\caption[ ]{Results for the NGC 3310 clusters from the 3DEF approach,
using the Starburst99 SSP models and adopting solar metallicity for
all clusters (cf. Figs. \ref{n3310ages.fig}d and
\ref{n3310masses.fig}d).}
{\scriptsize
\begin{center}
\begin{tabular}{cccccccccccc}
\hline
\hline
\multicolumn{1}{c}{ID} & \multicolumn{3}{c}{E$(B-V)$ (mag)} &&
\multicolumn{3}{c}{log( Age/yr )} && \multicolumn{3}{c}{log(
Mass/M$_\odot$ )}\\
\cline{2-4}\cline{6-8}\cline{10-12}
& \multicolumn{1}{c}{Min.} & \multicolumn{1}{c}{Best} &
\multicolumn{1}{c}{Max.} && \multicolumn{1}{c}{Min.} &
\multicolumn{1}{c}{Best} & \multicolumn{1}{c}{Max.} &&
\multicolumn{1}{c}{Min.} & \multicolumn{1}{c}{Best} &
\multicolumn{1}{c}{Max.} \\
\hline
G1-01 & 0.00 & 0.36 & 0.42 && 6.20 & 6.41 & 7.28 && 4.41 & 4.54 & 4.65 \\
G1-02 & 0.06 & 0.16 & 0.20 && 6.56 & 6.56 & 6.61 && 4.34 & 4.41 & 4.46 \\
G1-03 & 0.00 & 0.04 & 0.16 && 6.82 & 7.19 & 7.82 && 3.70 & 4.27 & 4.77 \\
G1-04 & 0.00 & 0.00 & 0.10 && 6.85 & 7.12 & 7.15 && 3.75 & 4.16 & 4.32 \\
G1-05 & 0.00 & 0.08 & 0.26 && 6.82 & 7.90 & 8.34 && 3.54 & 4.78 & 5.08 \\
G1-06 & 0.08 & 0.16 & 0.26 && 6.20 & 6.56 & 6.61 && 4.26 & 4.35 & 4.52 \\
G1-07 & 0.28 & 0.32 & 0.36 && 6.41 & 6.56 & 6.56 && 5.01 & 5.06 & 5.14 \\
G1-08 & 0.00 & 0.00 & 0.46 && 6.56 & 7.98 & 8.37 && 3.26 & 4.45 & 4.75 \\
G1-09 & 0.00 & 0.18 & 0.26 && 6.78 & 7.23 & 7.67 && 3.51 & 4.27 & 4.46 \\
G1-10 & 0.00 & 0.34 & 0.42 && 6.20 & 6.49 & 7.28 && 4.45 & 4.50 & 4.66 \\
G1-11 & 0.00 & 0.04 & 0.58 && 6.20 & 7.15 & 7.98 && 3.78 & 4.28 & 4.97 \\
G1-12 & 0.00 & 0.00 & 0.38 && 5.00 & 7.24 & 7.32 && 4.66 & 4.90 & 4.95 \\
G1-13 & 0.02 & 0.12 & 0.32 && 6.78 & 7.44 & 7.71 && 3.62 & 4.49 & 4.75 \\
G1-14 & 0.00 & 0.18 & 0.70 && 5.78 & 7.45 & 8.16 && 3.52 & 4.45 & 5.02 \\
G1-15 & 0.00 & 0.08 & 0.58 && 6.49 & 7.72 & 8.30 && 3.62 & 4.70 & 5.10 \\
G1-16 & 0.00 & 0.40 & 0.92 && 5.00 & 7.28 & 8.77 && 3.54 & 4.50 & 5.41 \\
G1-17 & 0.14 & 0.20 & 0.38 && 6.78 & 7.44 & 7.58 && 4.15 & 4.68 & 4.81 \\
\hline
\end{tabular}
\end{center}
}
\end{table*}

\begin{table*}
\caption[ ]{Results for the Antennae clusters from the
multi-dimensional SED fits, using the AnalySED approach with cluster
ages, masses, metallicities and extinction values as free parameters
(cf. Figs. \ref{n4038ages.fig}a and \ref{n4038masses.fig}a).}
{\scriptsize
\begin{center}
\begin{tabular}{ccccccccccrrcccr}
\hline
\hline
\multicolumn{1}{c}{ID} & \multicolumn{3}{c}{$Z$} &&
\multicolumn{3}{c}{E$(B-V)$ (mag)} && \multicolumn{3}{c}{Age ($\times
10^8$ yr)} && \multicolumn{3}{c}{Mass ($\times 10^5$ M$_\odot$)}\\
\cline{2-4}\cline{6-8}\cline{10-12}\cline{14-16}
& \multicolumn{1}{c}{Min.} & \multicolumn{1}{c}{Best} &
\multicolumn{1}{c}{Max.} && \multicolumn{1}{c}{Min.} &
\multicolumn{1}{c}{Best} & \multicolumn{1}{c}{Max.} &&
\multicolumn{1}{c}{Min.} & \multicolumn{1}{c}{Best} &
\multicolumn{1}{c}{Max.} && \multicolumn{1}{c}{Min.} &
\multicolumn{1}{c}{Best} & \multicolumn{1}{c}{Max.} \\
\hline
G2-01 & 0.0004 & 0.0500 & 0.0500 && 0.00 & 0.20 & 0.40 && 0.56 & 1.76 &10.50 && 0.55 & 1.63 & 2.47 \\
G2-02 & 0.0004 & 0.0080 & 0.0500 && 0.00 & 0.15 & 0.45 && 0.48 & 3.96 &11.70 && 0.54 & 1.84 & 2.54 \\
G2-03 & 0.0004 & 0.0500 & 0.0500 && 0.30 & 0.50 & 0.70 && 0.08 & 0.56 & 4.96 && 0.16 & 0.96 & 1.80 \\
G2-04 & 0.0200 & 0.0500 & 0.0500 && 0.15 & 0.35 & 0.45 && 0.08 & 0.12 & 0.40 && 0.38 & 1.25 & 2.38 \\
G2-05 & 0.0040 & 0.0500 & 0.0500 && 0.50 & 0.75 & 0.85 && 0.08 & 0.12 & 0.80 && 0.86 & 3.42 & 7.67 \\
G2-06 & 0.0004 & 0.0500 & 0.0500 && 0.00 & 0.15 & 0.25 && 0.60 & 1.64 & 8.04 && 0.50 & 1.36 & 1.90 \\
G2-07 & 0.0200 & 0.0500 & 0.0500 && 0.00 & 0.20 & 0.30 && 0.08 & 0.12 & 0.36 && 1.03 & 3.35 & 6.40 \\
G2-08 & 0.0200 & 0.0200 & 0.0500 && 0.40 & 0.45 & 0.70 && 0.08 & 0.16 & 0.40 && 0.54 & 0.94 & 3.37 \\
G2-09 & 0.0004 & 0.0200 & 0.0500 && 0.25 & 0.50 & 0.95 && 0.04 & 0.48 & 5.88 && 0.40 & 3.68 &10.90 \\
G2-10 & 0.0200 & 0.0500 & 0.0500 && 0.25 & 0.45 & 0.55 && 0.08 & 0.08 & 0.40 && 0.79 & 1.62 & 5.40 \\
G2-11 & 0.0004 & 0.0080 & 0.0500 && 0.55 & 0.95 & 1.00 && 0.08 & 0.16 & 5.12 && 0.17 & 1.51 & 5.12 \\
G2-12 & 0.0040 & 0.0200 & 0.0500 && 0.00 & 0.00 & 0.20 && 0.08 & 0.20 & 0.44 && 1.33 & 2.44 & 6.03 \\
G2-13 & 0.0004 & 0.0040 & 0.0500 && 0.50 & 0.95 & 1.00 && 0.08 & 0.16 &11.30 && 0.10 & 1.00 & 3.04 \\
G2-14 & 0.0040 & 0.0200 & 0.0500 && 0.70 & 0.75 & 1.00 && 0.08 & 0.08 & 0.64 && 0.28 & 0.34 & 3.91 \\
G2-15 & 0.0040 & 0.0200 & 0.0500 && 0.00 & 0.00 & 0.25 && 0.08 & 0.12 & 0.40 && 0.46 & 0.88 & 4.88 \\
G2-16 & 0.0040 & 0.0500 & 0.0500 && 0.05 & 0.25 & 0.45 && 0.08 & 0.32 & 2.52 && 0.28 & 2.20 & 3.97 \\
G2-17 & 0.0200 & 0.0500 & 0.0500 && 0.20 & 0.35 & 0.45 && 0.08 & 0.08 & 0.24 && 1.25 & 2.08 & 6.30 \\
G2-18 & 0.0004 & 0.0500 & 0.0500 && 0.00 & 0.20 & 0.40 && 0.60 & 2.08 &16.70 && 0.40 & 1.26 & 1.68 \\
G2-19 & 0.0004 & 0.0200 & 0.0500 && 0.40 & 0.65 & 0.95 && 0.08 & 0.32 & 3.60 && 0.12 & 1.07 & 2.99 \\
G2-20 & 0.0004 & 0.0200 & 0.0500 && 0.40 & 0.75 & 1.00 && 0.08 & 2.08 &125.~~~&&0.22 & 3.80 &15.50 \\
\hline
\end{tabular}
\end{center}
}
\end{table*}

\begin{table*}
\caption[ ]{Results for the Antennae clusters from the
multi-dimensional SED fits, using the AnalySED approach with cluster
ages, masses, and extinction values as free parameters, and adopting
solar metallicity for all clusters (cf. Figs. \ref{n4038ages.fig}b and
\ref{n4038masses.fig}b).}  {\scriptsize
\begin{center}
\begin{tabular}{cccccrrrcccc}
\hline
\hline
\multicolumn{1}{c}{ID} & \multicolumn{3}{c}{E$(B-V)$ (mag)} &&
\multicolumn{3}{c}{Age ($\times 10^8$ yr)} && \multicolumn{3}{c}{Mass
($\times 10^5$ M$_\odot$)}\\ 
\cline{2-4}\cline{6-8}\cline{10-12}
& \multicolumn{1}{c}{Min.} & \multicolumn{1}{c}{Best} &
\multicolumn{1}{c}{Max.} && \multicolumn{1}{c}{Min.} &
\multicolumn{1}{c}{Best} & \multicolumn{1}{c}{Max.} &&
\multicolumn{1}{c}{Min.} & \multicolumn{1}{c}{Best} &
\multicolumn{1}{c}{Max.} \\
\hline
G2-01 & 0.00 & 0.00 & 0.00 && 2.36 & 2.68 & 3.40 && 0.66 & 0.73 & 0.87 \\
G2-02 & 0.00 & 0.00 & 0.00 && 2.36 & 2.88 & 3.64 && 0.65 & 0.75 & 0.90 \\
G2-03 & 0.00 & 0.10 & 0.20 && 2.08 & 3.52 & 6.40 && 0.43 & 0.52 & 0.75 \\
G2-04 & 0.00 & 0.00 & 0.45 && 0.04 & 0.08 & 0.72 && 0.10 & 0.10 & 0.88 \\
G2-05 & 0.25 & 0.80 & 0.85 && 0.04 & 0.04 & 3.24 && 0.17 & 0.73 & 3.97 \\
G2-06 & 0.00 & 0.00 & 0.40 && 0.04 & 2.52 & 2.96 && 0.11 & 0.74 & 0.84 \\
G2-07 & 0.25 & 0.25 & 0.30 && 0.04 & 0.04 & 0.04 && 0.74 & 0.74 & 0.91 \\
G2-08 & 0.15 & 0.65 & 0.70 && 0.04 & 0.04 & 1.72 && 0.09 & 0.38 & 1.29 \\
G2-09 & 0.00 & 0.65 & 0.65 && 0.04 & 0.04 & 6.88 && 0.44 & 0.53 & 3.52 \\
G2-10 & 0.00 & 0.10 & 0.60 && 0.04 & 0.08 & 1.72 && 0.13 & 0.20 & 2.27 \\
G2-11 & 0.00 & 0.00 & 0.05 &&12.00 &20.80 &26.60 && 0.87 & 1.54 & 2.00 \\
G2-12 & 0.15 & 0.20 & 0.20 && 0.04 & 0.04 & 0.04 && 0.64 & 0.78 & 0.78 \\
G2-13 & 0.00 & 0.00 & 0.05 &&10.90 &18.00 &24.40 && 0.48 & 0.77 & 1.08 \\
G2-14 & 0.00 & 0.95 & 1.00 && 0.04 & 0.04 &34.60 && 0.09 & 0.37 & 3.50 \\
G2-15 & 0.20 & 0.25 & 0.25 && 0.04 & 0.04 & 0.04 && 0.51 & 0.63 & 0.63 \\
G2-16 & 0.00 & 0.00 & 0.40 && 0.04 & 0.64 & 0.68 && 0.31 & 1.00 & 1.22 \\
G2-17 & 0.00 & 0.00 & 0.50 && 0.04 & 0.08 & 1.28 && 0.25 & 0.25 & 2.96 \\
G2-18 & 0.00 & 0.00 & 0.00 && 2.40 & 2.68 & 3.36 && 0.46 & 0.50 & 0.59 \\
G2-19 & 0.00 & 0.75 & 0.85 && 0.04 & 0.04 & 9.84 && 0.05 & 0.16 & 1.21 \\
G2-20 & 0.00 & 0.00 & 0.20 && 8.60 &23.00 &31.00 && 0.91 & 1.55 & 2.08 \\
\hline
\end{tabular}
\end{center}
}
\end{table*}

\begin{table*}
\caption[ ]{Results for the Antennae clusters from the 3DEF approach,
using the Starburst99 SSP models and adopting solar metallicity for
all clusters (cf. Figs. \ref{n4038ages.fig}c and
\ref{n4038masses.fig}c).}
{\scriptsize
\begin{center}
\begin{tabular}{cccccccccccc}
\hline
\hline
\multicolumn{1}{c}{ID} & \multicolumn{3}{c}{E$(B-V)$ (mag)} &&
\multicolumn{3}{c}{log( Age/yr )} && \multicolumn{3}{c}{log(
Mass/M$_\odot$ )}\\
\cline{2-4}\cline{6-8}\cline{10-12}
& \multicolumn{1}{c}{Min.} & \multicolumn{1}{c}{Best} &
\multicolumn{1}{c}{Max.} && \multicolumn{1}{c}{Min.} &
\multicolumn{1}{c}{Best} & \multicolumn{1}{c}{Max.} &&
\multicolumn{1}{c}{Min.} & \multicolumn{1}{c}{Best} &
\multicolumn{1}{c}{Max.} \\
\hline
G2-01 & 0.00 & 0.00 & 0.08 && 8.24 & 8.50 & 8.69 && 4.49 & 4.62 & 4.76 \\
G2-02 & 0.00 & 0.00 & 0.09 && 8.24 & 8.50 & 8.71 && 4.50 & 4.61 & 4.78 \\
G2-03 & 0.00 & 0.26 & 0.50 && 6.81 & 8.33 & 9.00 && 3.22 & 4.44 & 4.70 \\
G2-04 & 0.00 & 0.36 & 0.51 && 6.48 & 6.78 & 8.10 && 3.52 & 4.10 & 4.77 \\
G2-05 & 0.27 & 0.61 & 0.84 && 6.55 & 6.81 & 8.44 && 3.71 & 4.03 & 5.25 \\
G2-06 & 0.00 & 0.00 & 0.04 && 8.11 & 8.33 & 8.52 && 4.37 & 4.53 & 4.65 \\
G2-07 & 0.00 & 0.21 & 0.37 && 6.40 & 6.78 & 7.36 && 4.18 & 4.59 & 4.89 \\
G2-08 & 0.11 & 0.27 & 0.72 && 6.55 & 6.85 & 8.11 && 3.47 & 3.60 & 4.84 \\
G2-09 & 0.00 & 0.00 & 0.46 && 6.81 & 8.98 & 9.00 && 3.89 & 5.32 & 5.41 \\
G2-10 & 0.00 & 0.48 & 0.62 && 6.48 & 6.78 & 8.15 && 3.74 & 4.40 & 5.10 \\
G2-11 & 0.21 & 0.28 & 0.75 && 6.81 & 8.99 & 9.00 && 3.50 & 4.89 & 4.99 \\
G2-12 & 0.00 & 0.10 & 0.25 && 6.48 & 6.78 & 6.81 && 4.26 & 4.55 & 4.70 \\
G2-13 & 0.18 & 0.26 & 0.50 && 8.40 & 9.00 & 9.00 && 4.41 & 4.64 & 4.72 \\
G2-14 & 0.25 & 0.32 & 0.93 && 6.78 & 9.00 & 9.00 && 3.44 & 5.09 & 5.17 \\
G2-15 & 0.00 & 0.21 & 0.33 && 6.48 & 6.55 & 6.81 && 4.10 & 4.30 & 4.60 \\
G2-16 & 0.00 & 0.16 & 0.39 && 6.55 & 6.81 & 8.37 && 3.66 & 3.83 & 5.06 \\
G2-17 & 0.00 & 0.42 & 0.58 && 6.40 & 6.74 & 8.04 && 3.96 & 4.65 & 5.24 \\
G2-18 & 0.00 & 0.00 & 0.07 && 8.28 & 8.50 & 8.67 && 4.33 & 4.45 & 4.59 \\
G2-19 & 0.07 & 0.59 & 0.66 && 6.81 & 6.81 & 9.00 && 3.17 & 3.43 & 4.89 \\
G2-20 & 0.23 & 0.31 & 0.54 && 8.48 & 9.00 & 9.00 && 4.71 & 4.90 & 4.98 \\
\hline
\end{tabular}
\end{center}
}
\end{table*}

\begin{table*}
\caption[ ]{Results for the Antennae clusters from the ``Sequential
O/IR'' approach, using the BC00 SSP models and metallicities ranging
from 0.4 to $2.5 Z_\odot$, as indicated in the column headings; the
$M/L_V$ ratios are given for solar metallicity
(cf. Figs. \ref{n4038ages.fig}d and \ref{n4038masses.fig}d).}
{\scriptsize
\begin{center}
\begin{tabular}{ccccccccc}
\hline
\hline
\multicolumn{1}{c}{ID} & \multicolumn{1}{c}{$A_V$} &
\multicolumn{3}{c}{log( Age/yr )} & \multicolumn{1}{c}{$M/L_V$} &
\multicolumn{3}{c}{log( Mass/M$_\odot$ )}\\
\cline{3-5}\cline{7-9}
& \multicolumn{1}{c}{(mag)} & \multicolumn{1}{c}{$(0.4 Z_\odot)$} &
\multicolumn{1}{c}{($Z_\odot$)} & \multicolumn{1}{c}{$(2.5 Z_\odot)$}
& \multicolumn{1}{c}{$(M/L_V)_\odot$} & \multicolumn{1}{c}{$(0.4
Z_\odot)$} & \multicolumn{1}{c}{($Z_\odot$)} &
\multicolumn{1}{c}{$(2.5 Z_\odot)$} \\
\hline
 G2-01 & 0.00 & 8.66 & 8.61 & 8.46 & 0.52 & 5.02 & 5.02 & 4.97 \\
 G2-02 & 0.00 & 8.71 & 8.61 & 8.51 & 0.52 & 5.04 & 5.02 & 5.01 \\
 G2-03 & 0.90 & 8.31 & 8.21 & 8.11 & 0.28 & 4.71 & 4.69 & 4.68 \\
 G2-04 & 0.14 & 6.82 & 7.22 & 7.56 & 0.06 & 3.92 & 4.50 & 4.84 \\
 G2-05 & 1.51 & 7.65 & 7.65 & 7.72 & 0.12 & 5.11 & 5.13 & 5.27 \\
 G2-06 & 0.00 & 8.56 & 8.41 & 8.36 & 0.36 & 4.96 & 4.89 & 4.89 \\
 G2-07 & 0.00 & 6.90 & 7.14 & 7.44 & 0.05 & 4.69 & 5.04 & 5.38 \\
 G2-08 & 0.49 & 6.82 & 7.04 & 8.01 & 0.04 & 3.87 & 4.20 & 4.98 \\
 G2-09 & 0.00 & 9.06 & 8.96 & 8.81 & 0.93 & 5.60 & 5.58 & 5.55 \\
 G2-10 & 0.50 & 6.82 & 7.44 & 7.56 & 0.09 & 4.27 & 4.97 & 5.14 \\
 G2-11 & 0.55 & 9.28 & 9.11 & 8.96 & 1.33 & 5.23 & 5.17 & 5.12 \\
 G2-12 & 0.00 & 6.76 & 6.76 & 6.68 & 0.02 & 4.71 & 4.68 & 4.66 \\
 G2-13 & 0.10 & 9.72 & 9.32 & 9.16 & 2.16 & 5.23 & 4.98 & 4.91 \\
 G2-14 & 0.00 & 9.76 & 9.34 & 9.16 & 2.26 & 5.64 & 5.38 & 5.29 \\
 G2-15 & 0.00 & 6.78 & 6.76 & 6.70 & 0.02 & 4.54 & 4.54 & 4.54 \\
 G2-16 & 0.22 & 7.70 & 7.81 & 7.81 & 0.15 & 4.99 & 5.08 & 5.16 \\
 G2-17 & 0.74 & 6.88 & 6.88 & 7.38 & 0.03 & 4.76 & 4.76 & 4.39 \\
 G2-18 & 0.00 & 8.66 & 8.61 & 8.51 & 0.52 & 4.86 & 4.87 & 4.87 \\
 G2-19 & 0.28 & 9.23 & 9.11 & 8.91 & 1.33 & 5.17 & 5.15 & 5.06 \\
 G2-20 & 0.54 & 9.48 & 9.26 & 9.01 & 1.88 & 5.39 & 5.30 & 5.15 \\
\hline
\end{tabular}
\end{center}
}
\end{table*}

\begin{table}
\caption[ ]{Results for the Antennae clusters from the {\it Q--Q} approach,
using the Kurth et al. (1999) SSP models and adopting solar metallicity for
all clusters (cf. Fig. \ref{n4038ages.fig}e).}
{\scriptsize
\begin{center}
\begin{tabular}{ccc}
\hline
\hline
\multicolumn{1}{c}{ID} & \multicolumn{1}{c}{$A_V$ (mag)} &
\multicolumn{1}{c}{log( Age/yr )}\\
\hline
G2-01 & 0.0 & 9.3 \\
G2-02 & 0.0 & 9.3 \\
G2-03 & 0.8 & 8.0 \\
G2-04 & 1.3 & 6.5 \\
G2-05 & 1.9 & 7.5 \\
G2-06 & 0.0 & 9.3 \\
G2-07 & 0.6 & 6.5 \\
G2-08 & 2.4 & 6.5 \\
G2-09 & 1.6 & 8.0 \\
G2-10 & 1.6 & 6.6 \\
G2-11 & 2.8 & 7.7 \\
G2-12 & 0.8 & 6.6 \\
G2-13 & 2.8 & 7.9 \\
G2-14 & 4.9 & 6.5 \\
G2-15 & 1.7 & 6.5 \\
G2-16 & 0.4 & 7.7 \\
G2-17 & 1.2 & 6.6 \\
G2-18 & 0.0 & 9.3 \\
G2-19 & 2.3 & 7.2 \\
G2-20 & 2.2 & 9.0 \\
\hline
\end{tabular}
\end{center}
}
\end{table}
\vfill\eject

\begin{table}
\caption[ ]{Results for the Antennae clusters from the BB+H$\alpha$
approach, using the BC00 SSP models and adopting solar metallicity for
all clusters (cf. Figs. \ref{n4038ages.fig}f and
\ref{n4038masses.fig}e).}
{\scriptsize
\begin{center}
\begin{tabular}{cccc}
\hline
\hline
\multicolumn{1}{c}{ID} & \multicolumn{1}{c}{E$(B-V)$ (mag)} &
\multicolumn{1}{c}{log( Age/yr )} & \multicolumn{1}{c}{Mass ($\times
10^5$ M$_\odot$)} \\
\hline
G2-01 & 0.00 & 8.56 &  0.96 \\
G2-02 & 0.00 & 8.61 &  1.03 \\
G2-03 & 0.38 & 8.01 &  0.49 \\
G2-04 & 0.34 & 6.79 &  0.21 \\
G2-05 & 0.76 & 6.78 &  0.45 \\
G2-06 & 0.00 & 8.41 &  0.77 \\
G2-07 & 0.42 & 6.46 & \  0.003 \\
G2-08 & 0.60 & 6.76 &  0.25 \\
G2-09 & 0.78 & 6.68 &  0.66 \\
G2-10 & 0.42 & 6.84 &  0.48 \\
G2-11 & 1.10 & 6.64 &  0.26 \\
G2-12 & 0.08 & 6.78 &  0.64 \\
G2-13 & 0.08 & 9.38 &  1.29 \\
G2-14 & 1.20 & 6.10 &  1.08 \\
G2-15 & 0.44 & 6.46 &  1.47 \\
G2-16 & 0.44 & 6.68 & \  0.007 \\
G2-17 & 0.30 & 6.86 &  0.66 \\
G2-18 & 0.00 & 8.61 &  0.74 \\
G2-19 & 0.96 & 6.58 &  0.20 \\
G2-20 & 0.24 & 9.16 &  1.92 \\
\hline
\end{tabular}
\end{center}
}
\end{table}

\end{document}